\documentclass[12pt]{article}

\usepackage{epsfig,latexsym,amsfonts,amsmath,amsthm,amssymb,amsbsy,multirow,slashed,wasysym,textcomp,dsfont,comment,mathtools,cancel,cite,diagbox,datetime,appendix,BOONDOX-calo}
\usepackage{tocloft}
\usepackage{bbold}
\usepackage{sidecap}
\usepackage{adjustbox}
\usepackage{pgfplots}
\usepackage{pgfkeys}
\usepackage{oplotsymbl}
\usepackage{tikzpagenodes}
\usepackage{adforn}
\tikzstyle{bag} = [align=center]
\usetikzlibrary{shapes,backgrounds,arrows.meta,patterns,decorations.markings,bending,positioning, decorations.pathmorphing,cd} 
\tikzset{snake it/.style={decorate, decoration=snake}}
\usepackage{adjustbox}
\usepackage[normalem]{ulem}
\usepackage[mathscr]{eucal}
\usepackage[version=3]{mhchem}
\usepackage{physics}

\usepackage[hidelinks]{hyperref}
\usepackage{subcaption}

 \newcommand{\badat}{\begin{alignedat}}
 \newcommand{\eadat}{\end{alignedat}}
 \newcommand\scalemath[2]{\scalebox{#1}{\mbox{\ensuremath{\displaystyle #2}}}}
 \def\be{\begin{equation}}
\def\ee{\end{equation}}

\usepackage{color}

\newcommand{\pink}[1]{\textcolor{\pink}{#1}}

\definecolor{dblue}{rgb}{0.2,0.50,0.80}

\usepackage[utf8]{inputenc}

\tikzset{snake it/.style={decorate, decoration=snake}}

\def\O{\mathcal{O}}

\def\bbR{\mathbb{R}}
\def\bbZ{\mathbb{Z}}

\def\bbE{\mathbb{E}}
\def\bbM{\mathbb{M}}

\def\RP{\mathbb{RP}}

\def\bz{{\bar z}}

\usepackage{tensor}
\usepackage{amsmath}
\usepackage{amssymb}
\usepackage{mathrsfs}
\usepackage[normalem]{ulem}
\usepackage{bbm} 
\usepackage{graphicx}
\usepackage{stmaryrd}
\usepackage{xcolor}
\usepackage{mathtools}
\usepackage{xfrac}

\renewcommand{\d}{\mathrm{d}}
\newcommand{\vphi}{\varphi}
\renewcommand{\bz}{\bar z}
\newcommand{\wt}[1]{\widetilde{#1}}

\def\la{\langle}
\def\ra{\rangle}

\def\bz{{\bar z}}

\def\im{{\rm i}}
\def\e{{\rm e}}

\def\scri{\mathscr I}
\def\sN{\mathscr{N}}

\newcommand{\til}[1]{\widetilde{#1}}
\def\d{\mathrm{d}}
\def\al{\alpha}

\def\CP{\mathbb{CP}}
\def\cN{\mathcal{N}}
\def\veps{\varepsilon}

\numberwithin{equation}{section} 
\pgfplotsset{compat=1.17} 
\begin{document}

\begin{titlepage}
\thispagestyle{empty}
  \begin{flushright}
  \end{flushright}
  \bigskip

  \begin{center}

                  \baselineskip=13pt {\LARGE \scshape{
                  Equating Extrapolate Dictionaries \\\vspace{0.4em} for Massless Scattering
            }
         }

      \vskip1cm

   \centerline{  Eivind J\o{}rstad$^{1,2}$, {Sabrina Pasterski}$^1$ and {Atul Sharma}$^3$   }

\bigskip\bigskip
 \bigskip\bigskip

 \centerline{\em${}^1$  
 Perimeter Institute for Theoretical Physics, }
 \vspace{0.2em}
 \centerline{\em Waterloo, ON N2L 2Y5, Canada}

\vspace{1em}

\centerline{\em${}^2$  
Dept. of Physics \& Astronomy, University of Waterloo,}
\vspace{0.2em}
\centerline{\em Waterloo, ON N2L 3G1, Canada}

\vspace{1em}

\centerline{\em${}^3$  
Center for the Fundamental Laws of Nature \& Black Hole Initiative,}
\vspace{0.2em}
\centerline{\em Harvard University, Cambridge MA 02138, USA}

\bigskip\bigskip

\end{center}

\begin{abstract}

We study features of celestial CFT correlation functions when the bulk theory is itself a CFT. We show that conformal inversions in the bulk map boost eigenstates to shadow transformed boost eigenstates. This is demonstrated explicitly for the wavefunctions of free massless scalars, and finds interesting applications to building extrapolate dictionaries. Because inversions exchange null infinity and the light cone of the origin, one finds a relation between the massless extrapolate dictionary -- involving correlators of operators inserted along null infinity  -- and the slice-by-slice extrapolate dictionary recently studied by Sleight and Taronna starting from the hyperbolic foliation of de Boer and Solodukhin. 
Namely, boundary correlators of Sleight and Taronna coincide with celestial amplitudes of shadow transformed boost eigenstates. These considerations are unified by lifting celestial correlators to the Einstein cylinder. This also sheds new light on the extraction of the $S$-matrix from the flat limit of AdS/CFT.

\end{abstract}

 \bigskip \bigskip \bigskip \bigskip

\end{titlepage}

\setcounter{tocdepth}{2}

\tableofcontents

\section{Introduction}

Understanding how the holographic principle extends from asymptotically AdS spacetimes to asymptotically flat spacetimes would give us a powerful toolkit to examine gravitational scattering. Over the years there have been various attempts at such a holographic description: from the work of BMS, Newman, Penrose, Ashtekar and others reconstructing bulk geometries from data at null infinity~\cite{Bondi:1962px,Sachs:1962wk,Sachs:1962zza, Newman:1976gc,Penrose:1976jq,Ashtekar:1981sf}, 
to attempts at using the AdS analog more directly: whether it be by extracting an $S$-matrix from the flat limit~\cite{Polchinski:1999ry,Penedones:2010ue} or, following de Boer and Solodukin, foliating Minkowski space with hyperboloids on which the (A)dS/CFT dictionary applies~\cite{deBoer:2003vf}.

The Celestial Holography program emerged from new
insights into the connection between symmetries and soft limits~\cite{Strominger:2017zoo}, which motivate reorganizing scattering~\cite{Pasterski:2016qvg,Pasterski:2017kqt} in terms of the infinite dimensional enhancements associated with the asymptotic symmetry group of asymptotically flat spacetimes. It proposes that quantum gravity in such spacetimes can be encoded in a codimension 2 dual living on the celestial sphere~\cite{Pasterski:2021rjz,Raclariu:2021zjz,Pasterski:2021raf}.
Here the bulk Lorentz group acts as the global conformal group and $S$-matrix elements are mapped to correlators of primaries in the Celestial CFT (CCFT) by going to a boost basis. The higher codimension of the dual is consistent with the expected number of degrees of freedom since each bulk field corresponds to a continuous spectrum of 2D primaries. 

The way these celestial primaries come about has different origins in the various routes to a flat space hologram we highlighted above. From the Bondi/Ashtekar~\cite{Bondi:1962px,Sachs:1962wk,Sachs:1962zza, Penrose:1976jq,Ashtekar:1981sf} point of view, we have a 4D extrapolate dictionary where the states corresponding to massless fields in the bulk are prepared by operators at null infinity. The boost basis is then a dimensional reduction of null infinity, where we are trading the null $u$ coordinate for a conformal weight. By contrast, in the hyperbolic foliation, we have a natural (A)dS$_3$/CFT$_2$ extrapolate dictionary on each leaf of the foliation, and we need to integrate over $|X|^2$~\cite{deBoer:2003vf,Cheung:2016iub}.\footnote{For concreteness we stick to 4D bulks for most of this paper though the same considerations hold for the massless extrapolate dictionary in a $D$-dimensional bulk and the (A)dS$_{d-1}$/CFT$_{d-2}$ slice dictionaries. See appendix~\ref{app:gendscalar}.} For brevity, in what follows we will refer to these as the 4D and 3D extrapolate dictionaries 
for 4D scattering. The 4D picture~\cite{He:2014laa,He:2020ifr, Pasterski:2021dqe} is what is implicitly used in the soft theorem =  Ward identity relations of~\cite{Strominger:2017zoo}. While the 3D picture lends itself to AdS/CFT re-interpretations of the flat space Ward identities~\cite{Cheung:2016iub,Ball:2019atb} and perturbative computations of amplitudes~\cite{Casali:2022fro, Iacobacci:2022yjo,Sleight:2023ojm}. 

While the physics is, of course, the same in both pictures, the `natural' expressions for the flat space amplitudes can look quite different. For example, following the 4D dictionary~\cite{Pasterski:2021dqe} one would write the $S$-matrix as correlators on null infinity 
\begin{equation}
\langle\mathcal{O}_{\Delta_1}(z_1,\bz_1)\cdots\mathcal{O}_{\Delta_n}(z_n,\bz_n) \rangle_{4D}  = \prod_{i}\lim_{r\rightarrow\infty}\int_{-\infty}^\infty \d 
\nu_i\, \nu_i^{-\Delta_i}\,\langle r\Phi(\nu_1,r,z_1,\bz_1)...r\Phi(\nu_n,r,z_n,\bz_n) \rangle\,
\label{eq:4DdictI}
\end{equation}
where $\nu=\{u,v\}$. Namely, we smear along the retarded time $u=t-r$ to prepare outgoing particles using operators at $\mathcal{I}^+$, and the advanced time $v=t+r$ to prepare incoming particles using operators at $\mathcal{I}^-$. Meanwhile~\cite{Sleight:2023ojm} presents expressions in terms of correlators on the light cone of the origin  
\begin{equation}
\langle\mathcal{O}_{\Delta_1}(Q_1)\cdots\mathcal{O}_{\Delta_n}(Q_n) \rangle_{3D}  = \prod_i \lim_{\hat{Y}_i\to Q_i}\int_0^\infty \d t_i\, t_i^{\Delta_i-1}\,\langle \Phi(t_1\hat{Y}_1)...\Phi(t_n\hat{Y}_n) \rangle\,,\label{eq:STdict}
\end{equation}
where $\hat{Y}_i$ denotes a point on the  `unit' (A)dS$_3$ slice $\hat{Y_i}^2=\pm1$, and $Q_i$ is a point on its boundary. Now null infinity and the lightcone of the origin are exchanged under a 4D conformal inversion. Here we show that for the special case of a conformally invariant 4D theory we can indeed recover the dictionary~\eqref{eq:STdict} starting from~\eqref{eq:4DdictI}. In doing so, we find interesting insights into the appearance of the shadow transform in these dictionaries, arising from the neat fact that conformal inversions in the bulk map boost eigenstates of weight $2-\Delta$ to shadow transforms of boost eigenstates of weight $\Delta$. 

The procedure for relating these two dictionaries also lends itself to more general expressions for $S$-matrix-like objects for general 4D CFTs. For example one can lift the construction back to the Einstein cylinder.  Then, by a different choice of conformal frame, one can analogously recast the $S$-matrix in terms of correlators on a light-sheet. Meanwhile, our constructions can more generally apply to weakly coupled scattering since there we are perturbing around the free CFT. While our goal is to relate to extrapolate dictionary expressions for the flat $S$-matrix, a nice side effect is that the difference between the 4D extrapolate and 3D expressions is seen to be closely tied to two different prescriptions for recovering the $S$-matrix in the flat limit: Hijano's Fourier transformed correlators~\cite{Hijano:2019qmi}, and Komatsu et al's position space limit~\cite{Komatsu:2020sag}.

 This paper is organized as follows. In section \ref{sec:geocon} we review the action of conformal inversions in 4D from the point of view of the $S^1\times S^3$ `celestial sphere' of embedding space and its $\mathbb{Z}_2$ quotient to the Poincar\'e patch. We then show how this inversion induces a shadow transformation on the 2D CCFT correlators in section \ref{sec:sheqinv}, first motivating it from a symmetry argument and then demonstrating it explicitly for the wavefunctions and field operators for a free massless scalar. Applying this to correlation functions in \ref{sec:twoextrap}, we are able to relate the 4D and 3D (slice-by-slice) extrapolate dictionaries used by~\cite{He:2014laa,He:2020ifr, Pasterski:2021dqe} and~\cite{deBoer:2003vf,Cheung:2016iub,Casali:2022fro,Sleight:2023ojm},
respectively, to describe celestial amplitudes.  In section \ref{sec:lift} we lift our construction back to correlation functions of operators on the Einstein cylinder. 
Then we close in section~\ref{sec:discussion} with a discussion of of our results. For concreteness we will focus on the massless scalar throughout. Some comments about conjugation by discrete symmetry related to the $\mathbb{Z}_2$ quotient of embedding space in section \ref{sec:geocon},  the higher-$D$ analog of the construction in section~\ref{sec:sheqinv}, and how the construction in~\ref{sec:lift} generalizes to other conformal frames, can be found in the appendix.


\section{Geometry of conformal inversions}\label{sec:geocon}

Denote Minkowski space by $\bbM=\bbR^{1,3}$, with Cartesian coordinates $X^\mu$, $\mu=0,1,2,3$. The Minkowski metric is taken to be $\eta_{\mu\nu} = \text{diag}(-1,1,1,1)$ and has signature $(1,3)$. Conformal inversions of $\bbM$ map the light cone of the origin to null infinity, which a priori lies outside $\bbM$. So, to study conformal transformations, it is useful to work with a conformal compactification of $\bbM$ that contains such boundaries. This is best introduced via an embedding space formalism.

\subsubsection*{The Poincar\'e patch on the Einstein cylinder}
Let $Z^I$ with index $I=-1,0,\dots,4$ be coordinates on embedding space $\bbR^{2,4}$. The light cone of the origin of $\bbR^{2,4}$ is cut out by
\be\label{lc0}
(Z^{-1})^2+(Z^0)^2 = (Z^1)^2+(Z^2)^2+(Z^3)^2+(Z^4)^2\,.
\ee
We remove the origin $Z^I=0$. The quotient of the remaining light cone by positive rescalings $Z^I\sim t Z^I$, $t\in\bbR_+$, provides a simple way to arrive at Penrose's conformal compactification of $\bbM$ \cite{Penrose:1962ij}.

After removing the origin, this light cone does not contain points at which $Z^{-1}=Z^0=0$. So in the quotient, we may choose the $Z^I$ to satisfy
\be\label{eq:scale}
(Z^{-1})^2+(Z^0)^2 = (Z^1)^2+(Z^2)^2+(Z^3)^2+(Z^4)^2 = 1
\ee
by performing the positive rescaling $Z^I\mapsto Z^I/\sqrt{(Z^{-1})^2+(Z^0)^2}$. The resulting space is topologically $S^1\times S^3$. We will refer to this as the celestial quadric. In the celestial quadric, the complement of the hyperplane $Z^{-1}+Z^4=0$ consists of two Poincar\'e patches, each a copy of $\bbM$. Up to positive rescalings, their embeddings are given by
\be\label{embed}
X^\mu\mapsto Z^I = \pm\left(\frac{1+|X|^2}{2},X^\mu,\frac{1-|X|^2}{2}\right)\,,
\ee
where $|X|^2 = \eta_{\mu\nu}X^\mu X^\nu$ is the Lorentzian norm.\footnote{Note that there are no absolute values being taken. We use this notation whenever we need to avoid confusion with the Cartesian component $X^2$.} 
The inverse map produces a 2-fold cover of $\bbM$
\be\label{embedinv}
X^\mu = \frac{Z^\mu}{Z^{-1}+Z^4}
\ee
as it does not distinguish between the Poincar\'e patches.  This geometry is a higher dimensional cousin of the covering discussed in~\cite{Guevara:2021tvr,Jorge-Diaz:2022dmy} for the celestial torus \cite{Atanasov:2021oyu}.

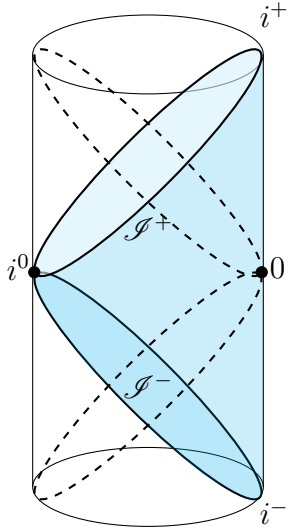
\begin{figure}[ht]
\centering
\vspace{-0.5em}
\begin{tikzpicture}[scale=2.1]
  \draw[] (0,-.64) ellipse (.73 and .25);
    \draw[] (0,2.1) ellipse (.73 and .25);
     \draw[thick,rotate=-45,fill=cyan, fill opacity=.1] (0,0) ellipse (1 and .18);
     \begin{scope}
  \clip (-.73,-1) rectangle (.73,3);
    \clip (3.4,-2) circle (5);
     \clip (0,0) circle (2.24);
     
    \clip   (-.65,0.5)--(-.7,0.6)--(-.725,0.73) --(-.63,1) -- (-.7,1.5) -- (1,2.3) -- (.72,-.7) --  (.65,-.727) -- (.32,-.55) -- (.05,-.3)--  (-.4,.15) --  cycle;
  \fill[cyan,fill opacity=.2] (3.4,3.4) circle (5);
\end{scope}

 \draw[rotate=45,thick,fill=white, fill opacity=.5] (1,1) ellipse (1 and .18);
        \draw[rotate=-45,dashed,thick] (-1,1) ellipse (1 and .18);
     \draw[rotate=45,dashed,thick] (0,0) ellipse (1 and .18);
     \draw[] (-.73,-.64) -- (-.73,2.1);
     \draw[] (.73,-.64) -- (.73,2.1);
     \node[] at (-.82,.75)  {$i^0$};
     \node[] at (.82,.75)  {$0$};
     \filldraw[black] (.72,.72) circle (1pt);
       \filldraw[black] (-.72,.72) circle (1pt);
     \node[] at (.8,2.35)  {$i^+$};
     \node[] at (.8,-0.8)  {$i^-$};
     \node[] at (0,0)  {$\scri^-$};
     \node[] at (0,1)  {$\scri^+$};
 \end{tikzpicture}

\caption{Poincar\'e patches on the Einstein cylinder. The celestial quadric $S^1\times S^3$ corresponds to identifying the top and bottom circles. The patch shaded in blue maps to the one in white under the map $Z^I\mapsto-Z^I$. An inversion exchanges the light cones of the origin (dashed) and spatial infinity (solid).}
\label{fig:einstein_cylinder}
\end{figure}

Let $\eta_{IJ}$ denote the flat metric on $\bbR^{2,4}$. The Minkowski metric $\d s^2=\eta_{\mu\nu}\d X^\mu\d X^\nu$ on $\bbM$ is obtained from the pullback of the weightless quadratic form
\be\label{ds2X}
\d s^2 = \frac{\eta_{IJ}\d Z^I\d Z^J}{(Z^{-1}+Z^4)^2}
\ee
by \eqref{embed}. Only the conformal structure of this metric extends to the boundary $Z^{-1}+Z^4=0$. The latter may be shown to consist of one copy of $\scri^+$ and $\scri^-$ shared between the two Poincar\'e patches, as in figure~\ref{fig:einstein_cylinder}, for the case where the circles at the top and bottom of the figure are identified. Fixing the scale to satisfy~\eqref{eq:scale}, 
one passes to the universal cover of $S^1\times S^3$ by introducing the covering map
\be\label{zm}
Z^I = 
\big(\cos\tau,\,\sin\tau,\,\sin\psi\sin\theta\cos\phi,\,\sin\psi\sin\theta\sin\phi,\,\sin\psi\cos\theta,\,\cos\psi\big)\,,
\ee
where $\tau\in\bbR$ and $(\psi,\theta,\phi)$ are spherical coordinates on $S^3$,
in particular \be
\phi\in[0,2\pi),~~~\psi,\theta\in[0,\pi).
\ee
The universal cover is the Einstein cylinder $\bbR\times S^3$ with its associated conformal structure. It consists of periodically repeating copies of conformally compactified Minkowski space. Each copy contains the usual null infinities $\scri^\pm$ and points at infinity $i^0,i^\pm$. Plugging \eqref{zm} into \eqref{embedinv}, we can map the Einstein cylinder to $\bbM$:
\be\label{xem}
X^\mu=\frac{(\sin\tau,\,\sin\psi\sin\theta\cos\phi,\,\sin\psi\sin\theta\sin\phi,\,\sin\psi\cos\theta)}{\cos\tau+\cos\psi}\,.
\ee
In particular, the $\mathbb{Z}_2$ quotient of the patches  is just $Z^I\mapsto-Z^I$ which is in $\mathrm{SO}(2,4)$ since we are in even dimensions. For the uplift to the cylinder we can use
\be\label{identified}
(\tau, \psi,\theta,\phi)\sim (\tau+\pi,\pi-\psi,\pi-\theta,\phi+\pi)
\ee
to relate each of the tiled Poincar\'e patches $\bbM$ in $\bbR\times S^3$.

\subsubsection*{Inversions on the Einstein cylinder}
The utility of the above construction is that the orientation preserving conformal group of $\bbM$ lifts to embedding space as the Lorentz group $\mathrm{SO}(2,4)$ that preserves the 6d light cone \eqref{lc0}. It also lifts to the Einstein cylinder as the universal cover of $\mathrm{SO}(2,4)$. Now we saw that the Poincar\'e patch is a $\mathbb{Z}_2$ quotient of the celestial quadric. Let's consider what an inversion looks like in the uplift to the celestial quadric and the Einstein cylinder. Indeed, it is only on embedding space or the Einstein cylinder that one can define a genuine action of conformal inversions, because they exchange the 4D light cone of the origin with null infinity, which is only contained in the conformal compactification. However, looking at figure~\ref{fig:einstein_cylinder}, one might be worried about the fact that exchanging the light cones of the origin and spatial infinity would imply the generators of $\scri^\pm$ map to points outside of the first Poincar\'e patch.

\begin{figure}[ht]
\centering
\vspace{-0.5em}
\begin{tikzpicture}[scale=2.1]
\definecolor{darkgreen}{rgb}{.0, 0.5, .1};
\draw[thick] (0,-1) -- (1,0) node[right]{$i^0$} -- (0,1);
\draw[thick] (0,-1) node[below]{$i^-$} -- (-1,0)  -- (0,1) node[above]{$i^+$};
\draw[thin,black!25!white] (-.5,.5)--(.5,-.5);
\draw[thin,black!25!white] (.5,.5)--(-.5,-.5);
\draw[thick, red] (.5,.5) to [bend left=45] (.5,-.5);
\draw[dashed,thick, red] (.5,.5) to [bend right=45] (.5,-.5);
\draw[thick, blue] (-.5,.5) to [bend left=45] (.5,.5);
\draw[dashed,thick, blue] (-.5,-.5) to [bend left=45] (.5,-.5);
\filldraw[blue] (.3,.63) circle (1pt);
\filldraw[blue] (-.3,-.38) circle (1pt);
\filldraw[red] (.63,.3) circle (1pt);
\filldraw[red] (.36,.3) circle (1pt);
\end{tikzpicture}
\caption{Inversion map $X^\mu\mapsto X^\mu/X^2 $ acting on a single Poincar\'e patch.  For reference, two sets of points on loci of fixed $X^2$ that are spacelike (red) and timelike (blue) separated from the origin are drawn before (solid) and after (dashed) the inversion. 
From the point of view of the $\mathbb{Z}_2$ uplift, $i^0$ maps to the origin. The red and blue points near null infinity on either side of $u=0$ map to points in different Poincar\'e patches.  Namely the red dot stays on the first Poincar\'e patch, while the blue dot maps to the second patch.  
}
\label{fig:inversion}
\end{figure}
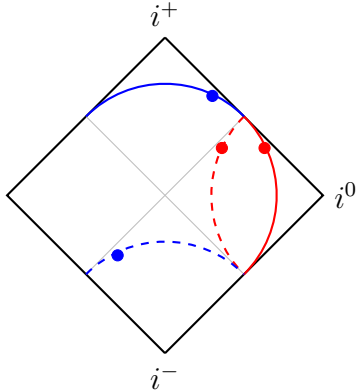

Let us visualize the lift of conformal inversions $X^\mu\mapsto X^\mu/|X|^2$. From the figure one can see that the origin and spatial infinity lie at antipodal points on the $S^3$. Indeed on can explicitly check that we can induce the standard inversion on $\mathbb{M}$ via a reflection of $\psi$ about $\psi=\pi/2$,
\be
X^\mu\mapsto  \frac{X^\mu }{|X|^2}~~~\Leftrightarrow~~~\psi\mapsto\pi-\psi\,.
\ee
This is equivalent to the reflection
\be\label{eq:ref}
Z^4\mapsto-Z^4
\ee
in embedding space. We thus see that the inversions of the Poincar\'e patch are in  $\mathbb{Z}_2=\mathrm{O}(2,4)/\mathrm{SO}(2,4)$.

Now we can always conjugate by an element of $\mathrm{SO}(2,4)$ to rotate the $Z^4$ reflection to one within the Lorentz group of the Poincar\'e patch $\mathrm{SO}(1,3)$. Here, we need to take some care with the fact that it's only proper orthochronous Lorentz transformations that are continuously connected to the identity $\mathrm{O}(1,3)/\mathrm{SO}^+(1,3)=\mathbb{Z}_2\times\mathbb{Z}_2$. Similarly, $\mathrm{SO}(2,4)$ itself has two components, with the Lie algebra exponentiating to $\mathrm{SO}^+(2,4)$. Upon conjugation by an element of $\mathrm{SO}(2,4)$ the reflection~\eqref{eq:ref} will still be in a spacelike direction. We thus expect the combination of spacetime inversions and parity of the Poincare patch to lift to an element of $\mathrm{SO}(2,4)$. Indeed
\be\label{eq:xProt}
X^\mu\mapsto P \frac{X^\mu }{|X|^2}P^{-1}~~~\Leftrightarrow~~~\tau\mapsto\tau+\pi.
\ee
From the point of view of the Einstein cylinder in figure~\ref{fig:einstein_cylinder} we see that composing a $\pi$ rotation in $\tau$, a parity transformation, and the inversion that reflects $Z^4$, maps us to the next Poincar\'e patch $Z^I\mapsto -Z^I$. We include some comments on how this perspective is useful for discussing the out states in the radial quantization picture of celestial CFT~\cite{Crawley:2021ivb} in appendix~\ref{app:conjugation}.

\subsubsection*{Compactification of $\mathbb{M}$ as a quadric in $\mathbb{RP}^5$}
For some calculations, it proves useful to work with an alternative compactification of $\bbM$ that only contains a single Poincar\'e patch \cite{Coxeter:1936} (see section 9.2 of \cite{Penrose:1986ca} for a review). This is obtained by quotienting the celestial quadric $S^1\times S^3$ by the residual $\bbZ_2$ action $Z^I\mapsto-Z^I$, reinterpreting the light cone $\eta_{IJ}Z^IZ^J=0$ as a quadric in $\RP^5$. Here, one treats the $Z^I$ as homogeneous coordinates, denoting by $[Z^I]$ their projective equivalence class. Minkowski space embeds into this quadric via
\be\label{embedZ2}
X^\mu\mapsto [Z^I] = \left[\frac{1+|X|^2}{2},X^\mu,\frac{1-|X|^2}{2}\right]\,,
\ee
which is simply the $\bbZ_2$ quotient of \eqref{embed}. 

The $\bbZ_2$ quotient also identifies $\scri^+$ with $\scri^-$ (namely, the start and endpoints of a null geodesic through the bulk get identified). So the hyperplane $Z^{-1}+Z^4=0$ now provides us with only a single copy of null infinity that we call $\scri$. Similarly, this compactification contains only a single point at infinity $[Z^I] = [1,0,0,0,0,-1]$. The conformal inversion $X^\mu\mapsto X^\mu/|X|^2$ is in fact cleanest in this presentation of the coordinates since, as we saw, it simply lifts to the linear action $Z^4\mapsto-Z^4$
up to projective rescalings. Because $|X|^2=Z^{-1}-Z^4$, we also see that this maps $\scri$ to the light cone of the origin and exchanges the point at infinity with the origin.

\subsubsection*{Inversions in Bondi coordinates}
Ultimately, we seek a picture of inversions in terms of the Bondi coordinates
\be\label{udef}
u = X^0 + r\,,\quad r = \sqrt{(X^1)^2+(X^2)^2+(X^3)^2}\,,\quad z = \frac{X^1+\im X^2}{r+X^3}\,.
\ee
Here, $u\in\bbR$, $r>0$, $z\in\CP^1$. In this parametrization, $|X|^2=-u(u+2r)$. So the light cone of the origin is the union of the sets
\be
\sN^+ = \{u=0\}\,,\qquad\sN^- = \{u=-2r\}\,
\ee
where $\sN^+$ is the future light cone while $\sN^-$ is the past light cone.  The embedding \eqref{embedZ2} may now be expressed as
\be
[Z^I] = \left[\frac{1-u(u+2r)}{2},u+r,\frac{r(z+\bz)}{1+|z|^2},\frac{\im r(\bz-z)}{1+|z|^2},\frac{r(1-|z|^2)}{1+|z|^2},\frac{1+u(u+2r)}{2}\right]\,.
\ee
Setting $u=0,-2r$ gives parametrizations of $\sN^\pm$:
\be\label{Npm}
    \sN^\pm\;:\quad[Z^I] = \left[1,\pm2r,\frac{2r(z+\bz)}{1+|z|^2},\frac{2\im r(\bz-z)}{1+|z|^2},\frac{2r(1-|z|^2)}{1+|z|^2},1\right]\,.
\ee
On the other hand, if one sets $R = 1/r$ and performs the projective rescaling $Z^I\mapsto RZ^I$, one can freely take the $r\to\infty$ limit by sending $R\to0$. The resulting points lie precisely in $Z^{-1}+Z^4=0$:
\be\label{scribondi}
\scri\;:\quad [Z^I] = \left[-u,1,\frac{z+\bz}{1+|z|^2},\frac{\im (\bz-z)}{1+|z|^2},\frac{1-|z|^2}{1+|z|^2},u\right]\,.
\ee
This is an explicit parametrization of $\scri$ in terms of Bondi coordinates. 

As seen above, performing the conformal inversion amounts to changing the sign of $Z^4$. The locus $u=0$ is the set of fixed points of this action and constitutes the celestial sphere of the origin. For $u\neq0$, acting with $Z^4\mapsto-Z^4$ and projectively rescaling by $-u^{-1}$ maps the points \eqref{scribondi} to points on the light cone of the origin:
\be\label{invscri}
[Z^I] = \left[1,-\frac{1}{u},-\frac{z+\bz}{u(1+|z|^2)},-\frac{\im (\bz-z)}{u(1+|z|^2)},-\frac{1-|z|^2}{u(1+|z|^2)},1\right]\,.
\ee
A comparison of \eqref{invscri} and \eqref{Npm} shows that if $u<0$, then $u\mapsto-1/2r$ whereas $z\mapsto z$. If $u>0$, one instead sees that $u\mapsto1/2r$ and $z\mapsto-1/\bz$. In other words, under inversions, cuts of $\scri$ at negative $u$ map to cuts of the future light cone $\sN^+$. And cuts of $\scri$ at positive $u$ map to cuts of the past light cone $\sN^-$ up to an antipodal map.\footnote{Note that in the antipodal matching of~\cite{Strominger:2017zoo}, one uses an antipodally related celestial sphere coordinate on past null infinity as compared to future null infinity. The amounts to using the projective coordinates for the future lightcone of $i^0$ to label the full generator extending along $\mathcal{I}^-$ and $\mathcal{I}^+$ in the conformal compactification. This lightcone maps to the lightcone of the origin under conformal inversions, consistent with the antipodal map appearing here.} In either case, one obtains the coordinate transformation
\be
r = \frac{1}{2|u|}\,.
\label{eq:invr}
\ee
This can be used to map integrals over $\scri$ to integrals over the light cone of the origin.



\section{2D shadow transform = 4D conformal inversion}\label{sec:sheqinv}

In this section, we show that shadow transforms of boost eigenstates may be obtained by performing bulk conformal inversions, thereby avoiding any integrals over $z,\bz$. This idea will play a central role in our understanding of the massless limits of various scattering dictionaries.
It works both at the level of wavefunctions as well as boundary operators defined using an extrapolate dictionary. As with the rest of this paper, we restrict attention to scalar states,\footnote{Massless scalars are automatically conformally coupled in flat space, as we discuss below.} but expect it to generalize to higher spins by the use of Penrose's conformally covariant zero rest mass equations \cite{Penrose:1965am} and from the explicit form of the conformal primary wavefunctions (CPW)~\cite{Pasterski:2017kqt,Pasterski:2021fjn}.

\subsection{Inversions commute with boosts}

Let us start with a symmetry based argument as motivation. Inversions are a conformal symmetry of the bulk, so they will map solutions of conformally covariant wave equations to new solutions. Before studying their action on such solutions, let's review the action of inversions at the level of the Lorentz algebra.

Let $X^\mu\mapsto(\Lambda X)^\mu\equiv\Lambda^\mu{}_\nu X^\nu$ denote a generic Lorentz transformation. Since $|\Lambda X|^2=|X|^2$, if we follow this by an inversion, we find the sequence of transformations
\be
X\mapsto\Lambda X\mapsto \frac{\Lambda X}{|\Lambda X|^2} = \frac{\Lambda X}{|X|^2}\,.
\ee
In the reverse order, if we first act by an inversion and then rotate by $\Lambda$, we find an identical end result:
\be
X\mapsto \frac{X}{|X|^2}\mapsto\Lambda\bigg(\frac{X}{|X|^2}\bigg) = \frac{\Lambda X}{|X|^2}\,.
\ee
Hence, inversions $I$ commute with the Lorentz generators $M_{\mu\nu}$,
\be\label{eq:IM}
I^{-1} M_{\mu\nu}I = M_{\mu\nu}\,.
\ee
A consequence of this will be that inversions will map boost eigenstates to boost eigenstates

We will see this explicitly for the wavefunctions in the next section, but let us first see this representation-theoretically. Let $L_0$ denote the left moving boost+rotation generator that stabilizes the point $(z,\bz)$ on the celestial sphere,\footnote{Explicitly, $L_0=\frac{1}{2}(J_3-iK_3)$ for $(z,\bz)=(0,0)$.} and let $|\Delta,z,\bz\ra$ be a boost eigenstate of a massless scalar field\footnote{We could just as well restore a non-trivial 2D spin $J$ here.}
\be
L_0|\Delta,z,\bz\ra = \frac{1}{2}\Delta|\Delta,z,\bz\ra\,.
\ee
Here, $\Delta$ denotes conformal weight and $z,\bz$ denote complex coordinates on the celestial sphere. Because $I^{-1}L_0I=L_0$, an inversion maps $|\Delta,z,\bz\ra$ to a new boost eigenstate of the same weight:
\be\label{eq:l0same}
\begin{split}
    L_0I|\Delta,z,\bz\ra &= IL_0|\Delta,z,\bz\ra\\
    &= \frac{1}{2}\Delta\,I|\Delta,z,\bz\ra\,.
\end{split}
\ee  
Moreover the primary condition is preserved
\be\label{eq:l1same}
\begin{split}
    L_1I|\Delta,z,\bz\ra 
    &= I L_1|\Delta,z,\bz\ra\,=0,
\end{split}
\ee
and similarly for ${\bar L}_1$. Meanwhile the translations are exchanged with special conformal transformations under inversions so, for example, we do not expect the additional Poincare primary conditions 
\be
P_{-\frac{1}{2},-\frac{1}{2}}|\Delta,z,\bz\rangle=|\Delta+1,z,\bz\rangle,~~~P_{\frac{1}{2},\frac{1}{2}}|\Delta,z,\bz\rangle=0,~~~P_{\pm\frac{1}{2},\mp\frac{1}{2}}|\Delta,z,\bz\rangle=0
\ee
to hold for $I|\Delta,z,\bz\rangle$, i.e. the two states $|\Delta,z,\bz\ra$ and $I|\Delta,z,\bz\ra$ will generically be distinct.

That is, if we include inversions, we find a degeneracy in boost eigenstates. Luckily, this lines up perfectly with the degeneracy found in \cite{Pasterski:2017kqt}. For every $\Delta$, there are two boost eigenstates: a conformal primary state of weight $\Delta$, and a different state obtained from shadow transforming a primary state of weight $2-\Delta$. For a general 2D CFT, the shadow transform~\cite{Ferrara:1972ay,Ferrara:1972uq,Ferrara:1972xe,Ferrara:1972kab} is an intertwiner between primaries with weight $(\Delta,J)$ and $(2-\Delta,-J)$.\footnote{More generally the shadow transform is an intertwiner between representations of weight $(\Delta,|J|)$ and $(d-\Delta,|J|)$ where $d=D-2$ is the dimension of the celestial sphere where the CFT is defined. We will illustrate the generic-$d$ scalar case in appendix~\ref{app:gendscalar}.} Its uniqueness up to normalization is discussed in~\cite{Simmons-Duffin:2012juh}.
Moreover the inversion squares to 1, while the shadow transform squares to a multiple of the identity, so this is consistent and fixes the normalization.

The Klein-Gordon equation in flat space happens to be conformally covariant, so it is meaningful to study the action of inversions on states procured by its solutions. Indeed, it is a feature of the celestial spectrum that we do have a state at $2-\Delta$ who's shadow image can map to the inversion but, again, this is consistent with the fact that our argument for the two states to be distinct relied on the translations which is what is requiring the principal series spectrum for a complete set of scattering states. From this symmetry argument, and feature of the spectrum, we thus see that
\be\label{eq:ishw}
{I}=\mathrm{sh}\circ \mathrm{W},
\ee where $I$ is an inversion, sh is the shadow transform and W is a Weyl reflection that sends $\Delta\mapsto 2-\Delta$, holds for the single particle states.
In the next section, we will show explicitly that if $\vphi(X)$ is the wavefunction of the state $|\Delta,z,\bz\ra$, then the inversion produces a new wavefunction $|X|^{-2}\vphi(X/|X|^2)$ of the same weight which indeed coincides with the shadow transform.

\subsection{Wavefunctions}\label{sec:wavefnc}
The most commonly used scattering wavefunctions for a massless scalar in flat space are the momentum eigenstates
\be\label{momstate}
    \vphi(X) = \e^{\im p\cdot X}\,,\qquad p_\mu p^\mu=0\,.
\ee
Here the null momentum $p_\mu$ can be parametrized as
\be\label{nullmom}
p_\mu = \veps\omega q_\mu(z,\bz)\,,\qquad q_\mu(z,\bz) = \frac{1}{2}\left(1+|z|^2,z+\bz,\im(\bz-z),1-|z|^2\right)\,,
\ee
where $\veps\in\{\pm1\}$ is the sign of the frequency, $\omega>0$ is the absolute value of the frequency, and $z,\bz$ are complex coordinates on the celestial sphere.

The inversion map
\be\label{eq:inv2}
X^\mu\mapsto \frac{X^\mu}{|X|^2}
\ee
is an orientation-reversing automorphism of compactified Minkowski space. Crucially, it induces a conformal transformation of the flat metric,
\be
|\d X|^2 \mapsto \frac{|\d X|^2}{|X|^4}\,.
\ee
It is an involution, so a second inversion maps the flat metric to itself. We can use this fact to generate new (local) solutions of the massless Klein-Gordon equation from older solutions.

The conformally covariant completion of the Klein-Gordon operator $\square$ associated to a metric $g$ is the conformal d'Alembertian, also known as the Yamabe operator,
\be
Y_g = \square - \frac{R}{6}\,,
\ee
where $R$ is the scalar curvature of $g$. Under a Weyl rescaling, it obeys the transformation law
\be\label{conflaw}
Y_{\Omega^2g}(\Omega^{-1}\varphi) = \Omega^{-3}Y_g\varphi\,.
\ee
So if $\varphi$ solves the free field equation $Y_g\varphi=0$, then $\Omega^{-1}\varphi$ solves the same but with respect to the conformally rescaled metric $\Omega^2g$. 

Taking $g = |\d X|^2$ and $\Omega=|X|^{-2}$, this gives
\be\label{yaminv}
Y_{\frac{|\d X|^2}{|X|^4}}\big(|X|^2\varphi(X)\big) = |X|^6\,Y_{|\d X|^2}\varphi(X)\,.
\ee
The flat space Yamabe operator $Y_{|\d X|^2}$ is just the flat space Klein-Gordon operator because the scalar curvature vanishes. Suppose we plug in an on-shell field $\varphi(X)$ that solves $\square\varphi(X) = 0$. Then the right hand side of \eqref{yaminv} vanishes, yielding
\be
Y_{\frac{|\d X|^2}{|X|^4}}\big(|X|^2\varphi(X)\big) = 0\,.
\ee
Performing the inversion $X^\mu\mapsto X^\mu/|X|^2$ on this yields our main identity:
\be
Y_{|\d X|^2}\left(\frac{1}{|X|^2}\,\varphi\bigg(\frac{X}{|X|^2}\bigg)\right) = 0\,.
\ee
This is telling us that any solution $\varphi(X)$ of the Klein-Gordon equation gives rise to another solution
\be
\wt{\vphi}(X) = \frac{1}{|X|^2}\,\varphi\bigg(\frac{X}{|X|^2}\bigg)\,.
\ee
For example, starting from the momentum eigenstate \eqref{momstate}, we are led to
\be\label{invmom}
\wt\varphi(X) = \frac{1}{|X|^2}\,\e^{\im p\cdot X/|X|^2}\,.
\ee
One can directly verify that this solves the Klein-Gordon equation everywhere except along the light cone of the origin where it becomes singular. This singularity may subsequently be tamed via an $\im\epsilon$ prescription.

Next, take $\varphi$ to be a boost eigenstate of weight $2-\Delta$,
\be
\vphi_{2-\Delta}(X) = \frac{1}{(\veps q\cdot X)^{2-\Delta}}\,,\label{eq:boosteigen}
\ee
with $q_\mu(z,\bz)$ as displayed in \eqref{nullmom}. Under inversion, this maps to the new solution
\be
\wt{\varphi}_{2-\Delta}(X) = \frac{(|X|^2)^{1-\Delta}}{(\veps q\cdot X)^{2-\Delta}}\,.
\label{eq:invwavefnc}
\ee
Up to a normalization factor 
\be
\cN_\Delta=\frac{\Gamma(\Delta)}{\pi\Gamma(\Delta-1)},
\ee
this is easily recognized to be the shadow transform of a boost eigenstate of weight $\Delta$ \cite{Pasterski:2017kqt}
\be
\wt{\varphi}_{2-\Delta}(X) = \cN_\Delta\,\mathbf{S}[\varphi_{\Delta}(X)]\,.
\label{eq:wavefncshadow}
\ee
Hence, at the level of wavefunctions of massless particles, the shadow transform is dutifully implemented by a conformal inversion. No integrals required! Parts of this idea were already noticed as part of the past work \cite{Pasterski:2021fjn,Brown:2022miw}. In particular, keeping track of the expected conformal factors for the fields is all one needs besides explicitly plugging the coordinate inversion~\eqref{eq:inv2} into the general spin wavefunctions in~\cite{Pasterski:2021fjn}.

It also follows that since a boost eigenstate is a Mellin transform of the momentum eigenstate \eqref{momstate}, its shadow transform is a Mellin transform of the inverted momentum eigenstate \eqref{invmom}. Explicitly, one finds
\be
    \frac{(|X|^2)^{1-\Delta}}{(\veps q\cdot X)^{2-\Delta}} = \frac{\im^{\Delta-2}}{\Gamma(2-\Delta)}\int_0^\infty\d\omega\,\omega^{1-\Delta}\,\frac{\e^{\im\veps\omega q\cdot X/|X|^2}}{|X|^2}\label{eq:planeinv}
\ee
where the Mellin integral is understood via an $\im\epsilon$ prescription. So the amplitudes of shadow transformed boost eigenstates will equal Mellin transforms of amplitudes of the inverted momentum eigenstates. 

\subsection{Operators}\label{sec:ops}

Now that we have seen how the 4D inversion image of the wavefunctions implements a shadow transform, we will now show that the relation~\eqref{eq:ishw} also holds when acting on the celestial operators. Namely, given a celestial operator $\mathcal{O}_\Delta$, we want to compute the inversion image $I \mathcal{O}_\Delta I^{-1}$ and will see that it is indeed related to the shadow operator $\widetilde{\cal O}_{\Delta}=\mathrm{sh}\circ{\cal O}_{2-\Delta}$.

We will start by considering the boundary operators obtained via the 4D extrapolate dictionary used in~\cite{ Pasterski:2021dqe}. As shown in~\cite{ Pasterski:2021dqe}, by pushing the Cauchy slice used to define the celestial operators in terms of conformal primary wavepackets to $\scri^+$ so that we can write the boost-basis $S$-matrix as
\begin{equation}
\langle out|S|in\rangle_{boost}=\langle\mathcal{O}_{\Delta_1}(z_1,\bz_1)\cdots\mathcal{O}_{\Delta_n}(z_n,\bz_n) \rangle_{4D}  
\label{eq:4Ddict}
\end{equation}
for the correlation function defined in~\eqref{eq:4Ddict}.
More generally, we have\footnote{For readability we are dropping an $\im\epsilon$ regulator in this expression. To select the annihilation operators that prepare the single particle out states we want to take $u^{-\Delta}\mapsto (u-\im \epsilon)^{-\Delta}$ in the integral kernel.} 
\begin{equation}
     \mathcal{O}_\Delta (z,\bar{z})\equiv \int^{\infty}_{-\infty} \d u\, u^{-\Delta}   \lim_{r\to\infty}\left[r^\delta \Phi(u,r,z,\bar{z})\right]\,,
     \label{eq:4Ddict0}
\end{equation}
with an analogous expression holding for an operator on $\scri^-$. These boundary operators are obtained by scaling bulk operators by an appropriate weight $r^\delta$ and taking $r\to\infty$ while keeping a null coordinate and angular coordinates fixed, closely analogous to the usual extrapolate dictionary in AdS$_4$/CFT$_3$~\cite{Banks:1998dd,Harlow:2011ke}. This $\delta$ depends on the 4D scaling dimension and spin of the bulk operator, and is distinct from the celestial boost weight $\Delta$, which arises from dimensionally reducing null infinity to the celestial sphere.
The appropriate integral kernel for preparing the boost eigenstates can be found by applying a combined Fourier transform to momentum space and Mellin transform to $\Delta$ space~\cite{Donnay:2022sdg}.  For simplicity, we consider a free massless scalar field in $3+1$ dimensions, for which $\delta=1$. Below we consider the bulk field operator $\Phi(X)$ and the ``inverted" operator $\til{\Phi}(X)=\frac{1}{|X|^2}\,\Phi\left(\frac{X}{|X|^2}\right)$ and show that the corresponding extrapolated operators are related by a shadow transform.\footnote{Note that we will rely on an expansion in creation and annihilation operators in the following. This is fine if we are interested in situations where the bulk theory can be treated perturbatively. A different, more general approach may be needed to treat theories whose interactions don't die off sufficiently fast enough at null infinity. Some comments to this effect are included in section~\ref{sec:lift}.}

Firstly, consider the bulk field operator $\Phi$(X). Recall that by applying the usual saddle point approximation, it is straightforward to show that the large $r$ limit takes the well-known form 
\begin{equation}\label{limrphi}
    \lim_{r\to\infty}r\Phi(u,r,z,\bar{z})=-\frac{\im}{8\pi^2}\int^\infty_0 \d\omega \, a(\omega,z,\bar{z})\, \e^{-\im\omega u}+\text{h.c.}\,,
\end{equation}
where $a(\omega,z,\bar{z})$ is the annihilation operator for a plane wave mode with frequency $\omega$ and null momentum pointing towards the point $(z,\bar{z})$ on the celestial sphere. Applying the smearing along null infinity according to \eqref{eq:4Ddict0} then gives
\begin{equation}
    \mathcal{O}_\Delta (z,\bar{z})= \int^{\infty}_{-\infty} \d u\, u^{-\Delta}   \lim_{r\to\infty}r\Phi \sim \int_0^\infty \d\omega\,\omega^{\Delta -1} a(\omega \hat{x}) + \text{h.c.},
\end{equation}
showing that the extrapolated operator is the Mellin transform of a sum of plane wave creation and annihilation operators with fixed angular direction of the momentum.

Secondly, consider the inverted operator $\til{\Phi}(X)$. Using again the plane wave mode expansion combined with \eqref{eq:wavefncshadow} and the Mellin inversion of \eqref{eq:planeinv}, one finds
\begin{align}
    \til{\Phi}(X)&=\int \frac{\d^3p}{(2\pi)^3 \omega}\, a(\vec{p})\, \frac{\e^{- \im p\cdot X/|X|^2}}{|X|^2} + \text{h.c.}\nonumber\\
    &=\int \frac{\d ^3 p}{(2\pi)^3 \omega}\, a(\vec{p})\int_{1+\im\bbR}\frac{\d\Delta'}{2\pi\im}\,\im^{2-\Delta'}\Gamma(2-\Delta')\,\omega ^{\Delta'-2}\,\cN_{\Delta'}\,\mathbf{S}\!\left[\vphi_{\Delta'}(X)\right] + \text{h.c.}\nonumber\\
    &= -\int\frac{ \d^2 z}{(2\pi)^3}\int_0^\infty \d\omega \int\frac{\d\Delta'}{2\pi\im}\int\frac{\d^2z'}{|z-z'|^{2-2\Delta'}} \frac{\cN_{\Delta'}\,\Gamma(2-\Delta')\,\omega^{\Delta'-1}}{(\im q'\cdot X)^{\Delta'}} \,a(\omega,z,\bar{z})+\text{h.c.}\nonumber\\
    &= -\int\frac{ \d^2 z\,\d^2z'}{(2\pi)^3}\int_0^\infty \d\omega \int\frac{\d\Delta'}{2\pi\im}\frac{\til\cN_{\Delta'}\,\omega^{\Delta'-1}a(\omega,z,\bar{z})}{|z-z'|^{2-2\Delta'}}\int_0^\infty\d\omega'\,\omega'^{\Delta'-1}\,\e^{-\im\omega'q'\cdot X} + \text{h.c.}\,,
\end{align}
where $q'\equiv q(z',\bz')$ and $\til\cN_{\Delta'} = \cN_{\Delta'}\Gamma(2-\Delta')/\Gamma(\Delta')$ is a normalization factor. 
In going from the second to the third line, we have substituted in the definition of the shadow transform. The fourth line is obtained by using the Mellin representation of $1/(\im q'\cdot X)^{\Delta'}$. 

If we now apply the smearing in the $u$ coordinate corresponding to a celestial operator of weight $2-\Delta$, we find 
\begin{multline}
    \til{\mathcal{O}}_{2-\Delta}(w,\bar{w})\equiv \lim_{r\to\infty}\int_{-\infty}^{\infty}\d u\,u^{\Delta-2} \,\til{\Phi}(X)\\
    =-\lim_{r\to\infty}\int\frac{\d\Delta'}{2\pi\im}\,\frac{\til\cN_{\Delta'}}{8\pi^3}\int_{\bbR^2_+}\d\omega \,\d\omega'\,\omega'\,^{\Delta'-\Delta} \,\omega ^{\Delta'-1}\int\frac{\d^2 z\,\d^2 z'}{|z-z'|^{2-2\Delta'}}\,a(\omega,z,\bar{z})\, e^{-\im \omega' r q'\cdot Q} + \text{h.c.}
\end{multline}
where we have used $X^\mu=u \hat{t}^\mu+rQ^\mu$, with $\hat t^\mu=(1,0,0,0)$ and $Q(w,\bar w)$ being the null vector pointing toward $w,\bar w$ on the celestial sphere. We can now use the saddle point approximation at large $r$, which gives a factor of $-\pi\omega'^{-1}\delta^2(z'-w)$ and kills the integral over $z'$,
\be
\til{\mathcal{O}}_{2-\Delta}(w,\bar{w}) = \int\frac{\d\Delta'}{2\pi\im}\,\frac{\til\cN_{\Delta'}}{8\pi^2}\int_{\bbR_+}\d\omega\,\omega^{\Delta'-1}\int_{\bbR_+}\frac{\d\omega'}{\omega'}\,\omega'^{\Delta'-\Delta}\int\frac{\d^2 z\,a(\omega,z,\bar{z})}{|z-w|^{2-2\Delta'}} + \text{h.c.}
\ee
The $\omega'$ integral then yields a delta function $2\pi\im\,\delta(\im(\Delta'-\Delta))$, interpreted suitably over the principal series~\cite{Donnay:2020guq}. This kills the integral over $\Delta'$. Thus we finally obtain
\begin{equation}
    \til{\mathcal{O}}_{2-\Delta}(w,\bar{w}) = \frac{\til\cN_{\Delta}}{8\pi^2}\int\frac{\d^2 z}{|w-z|^{2-2\Delta}}\int^\infty_0 \d\omega\,\omega^{\Delta-1}\,a(\omega,z,\bar{z})\sim\mathbf{S}\big[\mathcal{O}_{\Delta}\big](z,\bar{z}),
    \label{eq:shdwop}
\end{equation}
which shows that the inverted celestial operator of weight $2-\Delta$ is the shadow transform of the celestial operator of weight $\Delta$. This is the operator analogue of the relation between inverted and shadow wavefunctions described in \eqref{eq:wavefncshadow}.

As mentioned in the previous sections, inversions map $\scri^+$ to the light cone of the origin. So an insertion of $\til{\O}_{2-\Delta}(w,\bar w)$ along the light cone of the origin in the inverted spacetime is equivalent to the insertion of a shadow operator at null infinity. We expand upon this point in the next section.


\section{Relating holographic dictionaries}\label{sec:twoextrap}

Now there is an alternative (A)dS$_3$/CFT$_2$ inspired extrapolate dictionary for celestial amplitudes proposed in~\cite{Sleight:2023ojm}, which starts from the hyperbolic foliation of Minkowski space by Euclidean AdS$_3$ and Lorentzian dS$_3$ slices as in~\cite{deBoer:2003vf}. In this prescription, bulk operators are Mellin transformed over the hyperbolic radial coordinate $t=\sqrt{\mp X^2}$, leaving only the (A)dS coordinates, and pushed to the (A)dS boundary, viz. 
\begin{equation}
\langle\mathcal{O}_{\Delta_1}(Q_1)\cdots\mathcal{O}_{\Delta_n}(Q_n) \rangle_{3D}  = \prod_i \lim_{\hat{Y}_i\to Q_i}\int_0^\infty \d t_i\, t_i^{\Delta_i-1}\,\langle \Phi(t_1\hat{Y}_1)...\Phi(t_n\hat{Y}_n) \rangle\,.\label{eq:STdictII}
\end{equation}
Here, $\hat{Y}_i$ denotes a point on the 
`unit' (A)dS slice defined by $\hat{Y_i}^2=\mp1$, and $Q_i$ is a point on the boundary of that slice. 
In this section, we argue that this extrapolate dictionary is related to the 4D prescription of (\ref{eq:4DdictI}) by a shadow transform. The main idea is that by pushing $\hat{Y}_i$ to the boundary of the slice, the integrals in \eqref{eq:STdict} will run along the light cone of the origin (see figure~\ref{fig:dicts}), which, as we saw in section~\ref{sec:geocon}, maps to null infinity under an inversion. Furthermore, it turns out that the Mellin integrals in (\ref{eq:STdict}) transform in the right way under the inversion to land on the smearing in (\ref{eq:4Ddict}). We can then use the relation between shadows and inversions established in section~\ref{sec:sheqinv} to relate the two dictionaries.

\subsection{Wavefunctions from propagators }\label{sec:wfprop}
Before moving to the general case, it is interesting to discuss the relationship between the 3D and 4D dictionaries at the level of wave functions.  We will construct these wave functions from Feynman propagators, which form the building blocks for constructing more general (in particular, not necessarily 4D conformal) celestial amplitudes in perturbation theory. This will also provide a nice picture where the source free solutions we are used to can be viewed as having sources at the conformal boundary, outside the Poincar\'e patch.

A wavefunction can be obtained via the extrapolate dictionary in a natural way. Let us first note that the bulk Feynman propagator is a solution to the equation of motion up to a delta function source at the location of the second operator
\begin{equation}\label{eq:txy}
    \square_X \langle T\!\left\{ \Phi(X)\Phi(Y) \right\} \rangle = -\im\,\delta^{4}(X-Y)\,.
\end{equation}
If we extrapolate the point $Y$ to future (past) null infinity, the point $X$ will always be in the past (future) if it is in the bulk. Thus, by sending the source out to the boundary, we obtain a source free solution in the bulk. Here we compute these `extrapolate wavefunctions' in the two dictionaries and show that they are related by an inversion (and thus, by (\ref{eq:invwavefnc}), a shadow transform).

Below, there will be two distinct notions of $\im \epsilon$ regularization appearing. One is the regularization of the measures in the integral transforms in \eqref{eq:4DdictI} and \eqref{eq:STdict}, which is related to the usual $\im \epsilon$ prescription needed to define the conformal primary wavefunctions. The other regularization needed is that for the propagator. Indeed, recall that the position space Feynman propagator takes the form\begin{equation}
    \langle T\!\left\{ \Phi(X)\Phi(Y) \right\} \rangle  = \frac{\im}{4\pi^2}\frac{1}{(X-Y)^2}\,.
\end{equation}
Now we need to take some care with how to handle the singularity when $X$ and $Y$ are on each other's lightcone. Convergence of the integrals we encounter when Fourier transforming from momentum to position space require $(X-Y)^2$ to have a positive imaginary part. For the 3D extrapolate dictionary~\cite{Sleight:2023ojm}  the authors use
\begin{equation}\label{eq:pm}
    \langle T^{\pm}\!\left\{ \Phi(X)\Phi(Y) \right\} \rangle  = \frac{\im}{4\pi^2}\frac{1}{(X-Y)^2\pm \im \epsilon}\,
\end{equation}
for the position space (anti-)Feynman propagator, however this can be achieved by the standard Wick rotation from Euclidean signature $X^0-Y^0\mapsto (X^0-Y^0)(1-\im\epsilon)$, or also by  $X^0-Y^0\mapsto X^0-Y^0-\im\epsilon\,\rm{sign}(X^0-Y^0)$, which reduces to the $X^0 \mapsto X^0\pm\im\epsilon$ used in~\cite{Pasterski:2017kqt,Puhm:2019zbl} when $Y^0\rightarrow\pm\infty$ in the 4D extrapolate dictionary.  The fact that the inversion maps us between the two dictionaries will give us some perspective on the difference between the $\im\epsilon$ prescriptions in~\cite{Sleight:2023ojm} vs~\cite{Pasterski:2017kqt,Puhm:2019zbl}, which arise from time ordering the inverted correlators.

Finally, we note that some care should be taken when looking at the inversion image of our correlators and wavefunctions. In particular, since inversions do not preserve time ordering on the quotiented Einstein cylinder, we should expect the appearance of an unconventional "time ordering" when trying to relate the two dictionaries. To this end, we note that \eqref{eq:pm} can be rewritten as
\begin{equation}\label{eq:invprop}
      \left\langle \bar{T}^{\pm}\!\left\{ \tilde{\Phi}(X)\tilde{\Phi}(Y) \right\} \right\rangle  = \frac{\im}{4\pi^2}\frac{1}{(X-Y)^2\pm \text{sign}\left(X^2Y^2\right)\, \im \epsilon}\,
\end{equation}
 where we have made the coordinate change $X\mapsto X/X^2$ and the notation $\bar{T}$ indicates the unusual $\im\epsilon$-prescription above. This prescription will be important in picking out the right poles of the propagator when we consider the 3D dictionary below.

\paragraph{CPW from the 4D Extrapolate Dictionary}
Let us first consider the conformal primary wavefunction we extract in the 4D dictionary, extrapolating to $\scri^+$ for concreteness. 
We can decompose $Y$ into
\begin{equation}
    Y = u \hat{t} +r \hat{q}(z,\bar{z}),
    \label{eq:parY}
\end{equation}
where $\hat{t}=(1,0,0,0)$ and $\hat{q}(z,\bar{z})$ is a null vector pointing towards the point $(z,\bar{z})$ on the celestial sphere defined by
\begin{equation}
    \hat{q}(z,\bar{z})\equiv\frac{2}{1+|z|^2}q(z,\bar{z}),
\end{equation}
with $q(z,\bar{z})$ given by \eqref{nullmom}.
 Just as we obtained the boundary operator in \eqref{eq:4Ddict}, one can obtain a boost eigenstate wavefunction by integrating this propagator in $u$ and extrapolating the result to $r\to\infty$.
The $u$ integral can be performed as a contour integral. The propagator \eqref{eq:pm} has two poles, namely
\begin{equation}
        u=(r \hat{q}(z,\bar{z})-X)\cdot\hat{t} \pm \sqrt{((r \hat{q}(z,\bar{z})-X)\cdot\hat{t})^2-2r X\cdot \hat{q}(z,\bar{z})+X^2\,\pm\im\epsilon}\,.
\end{equation}
In the large $r$-limit the poles are located at 
\begin{equation}
    \begin{split}
        u_1=& -X\cdot \hat{q}(z,\bar{z})\pm \im\epsilon +O(r^{-1}),\\
        u_2=& -2r \mp \im \epsilon + O(r^0),
        \label{eq:poles4D}
    \end{split}
\end{equation}
where we have absorbed some powers of $r$ in $\epsilon$, only keeping track of the sign. The $\im\epsilon$'s aboveindicate a choice of integration contour in $u$. The corresponding residues of interest are
\begin{equation}
    \begin{split}
        &\text{Res}\left[u^{-\Delta}  \langle T^{}\!\left\{ \Phi(X)\Phi(Y) \right\} \rangle\right]\big|_{u=u_1}=-\frac{\im}{8\pi^2 r}\frac{1}{(-\hat{q}\cdot X)^\Delta}+O(r^{-2}),\\
          &     \text{Res}\left[  u^{-\Delta} \langle T^{}\!\left\{ \Phi(X)\Phi(Y) \right\} \rangle\right]\big|_{u=u_2}=\frac{\im}{8\pi^2r}\,\frac{1}{(-2 r) ^{\Delta} }+O(r^{-3}),
    \end{split}
\end{equation}
where we recognize the first pole as the usual conformal primary wavefunction up to normalization. 

Now, the integral measure in (\ref{eq:4Ddict}) has a branch point at $u=0$, which needs to be dealt with. The usual prescription dictates that we choose $u\to u-\im\epsilon'$ for an outgoing operator, which pushes the branch point to the positive imaginary axis. Thus, we have to close the contour in the negative half plane, which gives
\begin{equation}
   \int^{\infty}_{-\infty} \d u \, (u - \im\epsilon')^{-\Delta} \lim_{r\to\infty} r\langle T^-\!\left\{ \Phi(X)\Phi(Y) \right\} \rangle  = -\frac{\im}{8\pi^2} \frac{1}{(-\hat{q}\cdot X-\im\epsilon')^\Delta}\sim \vphi_{\Delta}^+(X), 
\end{equation}
which is the usual outgoing conformal primary wavefunction. Note that $\epsilon'$ is distinct from the $\epsilon$ in \eqref{eq:poles4D}, which indicates the pole prescription. Also note \begin{equation}
   \int^{\infty}_{-\infty} \d u \, (u - \im\epsilon')^{-\Delta} \lim_{r\to\infty} r\langle T^+\!\left\{ \Phi(X)\Phi(Y) \right\} \rangle  = 0, 
\end{equation}
so the $\im\epsilon'$ for the $u$ integral indeed plays the role of selecting the appropriate sign frequency solutions as in~\cite{Pasterski:2021dqe}. Meanwhile, the second pole is of little interest here, as it vanishes at large $r$. The final result is indeed consistent with our discussion above that when we take $Y^0\rightarrow\pm\infty$ the regulator in~\eqref{eq:pm} is equivalent to the analytic continuation of $X^0\mapsto X^0\pm \im\epsilon$ used in~\cite{Pasterski:2017kqt}, and should match to the wavefunctions there.

\paragraph{CPW from the 3D Extrapolate Dictionary}
Let us now consider the 3D dictionary. There are two ways to take a past/future extrapolate limit in this case: either as a boundary limit on an Euclidean AdS$_3$ slice or on a dS$_3$ slice. We can make the following parameterizations
\begin{equation}\label{AdS}
    \begin{split}
       & \hat{Y}_{\text{AdS}}(\rho,z,\bar{z})=\rho \hat{q}(z,\bar{z})+\left(\sqrt{1+\rho^2}-\rho\right)\hat{t}\sim \rho\,\hat{q}(z,\bar{z})+\frac{1}{2 \rho}\hat{t}+O(\rho^{-2})\\
        & \hat{Y}_{\text{dS}}(\rho,z,\bar{z})=\rho q(z,\bar{z})+\left(\sqrt{\rho^2-1}-\rho\right)\hat{t}\sim \rho\,\hat{q}(z,\bar{z})-\frac{1}{2 \rho}\hat{t}+O(\rho^{-2}),
        \end{split}
\end{equation}
corresponding to the unit hyperboloids with $\hat{Y}_{\text{AdS}}^2=-1$ and $\hat{Y}_{\text{dS}}^2=1$. Note that the future extrapolate limit (i.e. taking the operator to $\scri^+$) is $\rho\to\infty$ for both slicings. In this limit the two lines of~\eqref{AdS} are approaching the null vector $q(z,\bar{z})$ from different sides of the light cone.  For the moment we will leave the choice of slicing ambiguous and simply write $\hat{Y}$.

 Here, we will define the 3D dictionary wavefunction with the the time ordering of \eqref{eq:invprop} in anticipation of the fact that the dictionaries will be related by bulk inversions, giving
\begin{equation}\label{eq:3Dwave}
    \phi_{\Delta}^\pm(X,Q)\equiv \lim_{\hat{Y}\to Q}\int_0^\infty dt\,t^{\Delta-1}\langle\bar{T}^{\mp}\{\Phi(X)\Phi(t\hat{Y}_{\text{}})\}\rangle.
\end{equation}
Some discussion of what would happen for a different choice of time ordering will follow towards the end of this section. Our goal here is not to calculate wavefunction \eqref{eq:3Dwave} explicitly, but rather to show that a certain natural linear combination of these 3D wavefunctions gives the inversion of the 4D wavefunction. 

 To this end, we are going to perform a similar residue calculation as for the 4D case. As a function of $t$ the propagator has two poles, namely
\begin{equation}
    t=\frac{1}{\hat{Y}^2}\left(X\cdot \hat{Y}\pm \sqrt{(X\cdot\hat{Y})^2 - \hat{Y}^2X^2\pm\text{sign}(X^2)\,\im\epsilon}\right).
\end{equation}
 In the large $\rho$ limit we get  
\begin{equation}
    \begin{split}
        t_1=& \frac{X^2}{2\rho\,(X\cdot \hat{q})}(1\pm \text{sign}(\hat{Y}^2)\,\im \epsilon)+O(\rho^{-1}),\\
        t_2=&\hat{Y}^2\,2\rho \, (X\cdot \hat{q})(1\mp \text{sign}(X^2)\,\im\epsilon)+O(\rho^{-1}),
    \end{split}
    \label{eq:polest}
\end{equation}
where as before any pre-factors of $\im\epsilon$ only matter in so far as they contribute an overall sign, which determines the integration contour. The corresponding residues are
\begin{equation}
\begin{split}
     &\text{Res}\left[  t^{\Delta-1}\langle T\big\{ \Phi(X)\Phi(t \hat{Y}) \big\} \rangle\right]\big|_{t=t_1}= -(2\rho)^{-\Delta}\frac{\im}{4\pi^2}\frac{\left(X^2\right)^{\Delta-1}}{(X\cdot \hat{q})^{\Delta}}+O(\rho^{-2}),\\
      &     \text{Res}\left[ t^{\Delta-1} \langle T\big\{ \Phi(X)\Phi(t\hat{Y}) \big\} \rangle\right]\big|_{t=t_2}=(2 \rho)^{\Delta-2} \frac{\im}{4\pi^2}  (\hat{Y}^2
      \,X\cdot \hat{q} )^{\Delta-2} + O(\rho^{-2}).
\end{split}\label{eq:resq}
\end{equation}
Note that the first pole is the inverted wavefunction (\ref{eq:invwavefnc}), and that the two poles are shadow transforms of one another (in both cases, up to prefactors involving $\rho$). 

Note also that both poles oscillate with $\rho$, which might seem like it would cause an issue when taking the  boundary limit $\rho\to\infty$. To understand how to handle this, let us unpack what it it really means to take $\hat{Y}\rightarrow Q$ in~\eqref{eq:STdictII}. When we are looking at $n$-point functions on the celestial sphere, the correlators are covariant under an overall rescaling of the null vector $Q_i\rightarrow \lambda Q_i$. However, the components of the null vector we get by taking $\hat{Y}$ on the unit hyperboloid to the boundary will diverge as $\rho\rightarrow \infty$. If we want to compare our wavefunctions to the ones with $q$ normalized as in~\eqref{nullmom}, we can instead rewrite~\eqref{eq:STdictII} as the following limit\footnote{Note that by holding $t_i$ fixed as we send $\rho\rightarrow\infty$ we are indeed integrating over a locus that approaches the light cone of the origin $X^2=-\frac{t^2}{\rho^2}=0$.} 
\begin{equation}
\langle\mathcal{O}_{\Delta_1}(q_1)\cdots\mathcal{O}_{\Delta_n}(q_n) \rangle_{3D}  = \prod_i \lim_{\rho\to \infty}
\int_0^\infty \d t_i\, t_i^{\Delta_i-1}\,\langle \Phi(t_1\hat{Y}_1/\rho)...\Phi(t_n\hat{Y}_n/\rho) \rangle\,.\label{eq:STdict2}
\end{equation}
Rescaling the integration variables to $t'_i=t_i/\rho$ this becomes
\begin{equation}
\langle\mathcal{O}_{\Delta_1}(q_1)\cdots\mathcal{O}_{\Delta_n}(q_n) \rangle_{3D}  = \prod_i \lim_
{\rho\to \infty}\int_0^\infty \d t_i'\, t_i'^{\Delta_i-1}\,\langle \rho^{\Delta_1}\Phi(t'_1\hat{Y}_1)...\rho^{\Delta_n}\Phi(t'_n\hat{Y}_n) \rangle\,,\label{eq:STdict2}
\end{equation}
as expected from our statement above about homogeneity in the $Q_i$. Upon doing this rescaling we see that the first line of~\eqref{eq:resq} scales like $\rho^0$ while the second scales like $\rho^{2\Delta-2}$. For $\Delta$ on the principal series the latter oscillates like $\rho^{i2\lambda}$. Note that we actually encounter similar behavior when we look at the large-$r$ expansion of the conformal primary wavefunctions~\cite{Donnay:2022sdg}. The term that is picked out by the saddle point approximation for the scalar field is a contact term that goes like $O(r^{-1})$ while the next term goes like $O(r^{-\Delta})$, so the extrapolate dictionary~\eqref{eq:4Ddict} is similarly selecting only the term that isn't oscillating in the corresponding limit. The coefficients of the two powers in $r$ are also related by a shadow transform of the reference direction. As such here we will similarly interpret the large-$\rho$ limit as selecting out the first line of~\eqref{eq:resq}.

Let us now extract the desired CPW. We want to add two 3D-dictionary wavefunctions in a way that essentially extends the integral in~\eqref{eq:3Dwave} to the whole real axis, and thus picks up one of the poles in \eqref{eq:polest}.  To extend the integral, we have to deal with two subtleties. Firstly, we need to pick a branch, since the measure $t^\Delta$ has a branch cut along the negative real axis. As we shall see in the next section, a natural regularization of the measure that handles this is $t^{\Delta-1}\to (1/t-\im\epsilon')^{-\Delta}/t$ for for an outgoing operator. This moves the branch point at infinity to $t=-\im/\epsilon$ while keeping the branch point at the origin fixed. We can then simply deform the integration contour by a small semicircle around the origin to continue the integral to the negative real axis.

The second subtlety is that the sign of the imaginary part of \eqref{eq:polest} depends on the sign of $X^2$ or $X\cdot q$. This fact is true both for the time ordering $\bar{T}^\pm$ and $T^{\pm}$. Thus, a naive integral over the $t$-axis will pick up different poles depending on the value of $X$. If we wish to pick out a definite pole, we can make the following combination (leaving the aforementioned regularization of the measure implicit)
\begin{equation}
\begin{split}
     e^{i\pi\Delta}\phi&
     _{\Delta,\text{dS}}^\pm(X,Q)- \phi_{\Delta,\text{AdS}}^\pm(X,-Q)\\
     &=\lim_{\rho\to\infty}\int^{\infty}_{0}dt\,t^{\Delta-1}\left[e^{i\pi\Delta}\langle \bar{T}^\mp\{ \Phi(X)\Phi(t \hat{Y}_{\text{dS}}/\rho) \} \rangle -\langle \bar{T}^\mp\{ \Phi(X)\Phi(-t \hat{Y}_{\text{AdS}}/\rho) \} \rangle \right]
\end{split}
\label{eq:extended}
\end{equation}
This combination of AdS and dS wavefunctions is what we would expect from our discussion of inversion images of the 4D dictionary and from figure~\ref{fig:inversion}. In the large $\rho$ limit, this becomes an integral along a null ray parallel to $\hat{q}(z,\bar{z})$. The pole location is naturally the same as in \eqref{eq:polest}, except for the fact that we have replaced factors of $\hat{Y}^2$ by $\text{sign}(t)$. Thus, one can see that the sign of the imaginary part of the pole position no longer depends on $X$. 

Finally, we evaluate \eqref{eq:extended} using the residue theorem. With the aforementioned regularization of the measure, which puts the branch cut in the negative imaginary half plane, we are forced to close the contour in the positive imaginary half plane. 
Thus, we get
\begin{equation}
     e^{\im\pi\Delta}\phi^{+}_{\Delta,\text{dS}}(X,Q)- \phi^{+}_{\Delta,\text{AdS}}(X,-Q)=-2^{-\Delta}\frac{\im}{4\pi^2}\frac{\left(X^2\right)^{\Delta-1}}{(-X\cdot q-\text{sign}(X^2)\im\epsilon')^{\Delta}},
     \label{eq:invSTwf}
\end{equation}
which (up to a power of $2$) is precisely the regularized inverted conformal primary wavefunction of \eqref{eq:invwavefnc}.\footnote{The singular behaviour in the numerator can be regularized by $X^2\to X^2 +\text{sign}(X\cdot q)\im\epsilon$, coming from the imaginary part in \eqref{eq:polest}. } Furthermore, by the argument below \eqref{eq:STdict2}, we have
\begin{equation}
     \phi^{-}_{\Delta,\text{dS}}(X,Q)-e^{\im\pi\Delta} \phi^{-}_{\Delta,\text{AdS}}(X,-Q)=0.
     \label{eq:invSTwf}
\end{equation}

Thus, we see that the 3D dictionary, when the smearing is extended to both the past and future light cone of the origin, gives us an inverted conformal primary wavefunction. 
This is due to the fact that the light cone of the origin maps to null infinity under inversions. Note that the massless limit of the wavefunctions in~\cite{Sleight:2023ojm} have a different $\im\epsilon$ prescription for avoiding the singularity at the lightcone $X^2=0$ coming from their regularization of the Feynman propagator.  From our statement that inversions exchange the light cone of the origin and $i^0$, and the manner in which we can view the free plane waves as having sources at null infinity via~\eqref{eq:txy} with $Y$ sent to $\scri^+$, we can now understand why removing the regulator would result in a source at the intersection of the light cone and the hyperplane $q\cdot X=0$ \cite{Pasterski:2020pdk}.

\subsection{Operators and correlation functions}
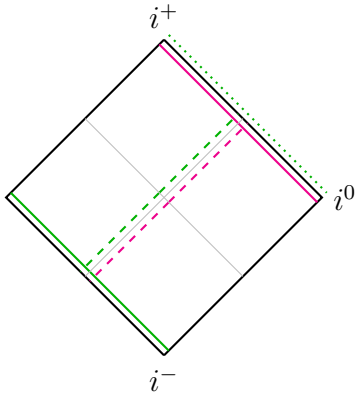
\begin{figure}[ht]
\centering
\vspace{-0.5em}
\begin{tikzpicture}[scale=2.1]
\definecolor{darkgreen}{rgb}{.0, 0.5, .1};
\draw[thick] (0,-1) -- (1,0) node[right]{$i^0$} -- (0,1);
\draw[thick] (0,-1) node[below]{$i^-$} -- (-1,0)  -- (0,1) node[above]{$i^+$};
\draw[thick,magenta] (0-.03,1-.03)--(1-.03,0-.03);
\draw[thick,dashed,black!30!green] (0-.03,0+.03)--(.5-.03,.5+.03);
\draw[thick,dashed,magenta] (0+.03,0-.03)--(.5+.03,.5-.03);
\draw[thick,dashed,black!30!green] (0-.03,0+.03)--(-.5-.03,-.5+.03);
\draw[thick,dashed,magenta] (0+.03,0-.03)--(-.5+.03,-.5-.03);
\draw[thick, dotted,black!30!green] (0+.03,1+.03)--(1+.03,0+.03);
\draw[thick,black!30!green] (-1+.03,0+.03)--(0+.03,-1+.03);
\draw[thin,black!25!white] (-.5,.5)--(.5,-.5);
\draw[thin,black!25!white] (.5,.5)--(-.5,-.5);
\end{tikzpicture}
\caption{Inversion image (dashed) of null generators being pushed to $\scri^+$ (magenta) and $\scri^-$ (green). However one should take care to note that the contributions near the past and future lightcones are the images of different Poincar\'e patches on the Einstein cylinder. Indeed, in the $\mathbb{Z}_2$ uplift, the dotted and solid green lines would map to the same locus upon quotienting to a single Poincar\'e patch. 
}
\label{fig:inversionscri}
\end{figure}

Let us now argue more generally that the 3D slice dictionary is related to the 4D dictionary by a shadow transform.\footnote{Note that the massive wavefunctions considered in~\cite{Sleight:2023ojm} before they restrict to the 4D CFT examples are shadow symmetric. Namely the shadow transform simply implements a Weyl reflection~\cite{Pasterski:2017kqt}. As such they would not have encountered this distinction between extracting the celestial and shadow celestial amplitudes.} First, let us consider the relationship between the dictionaries on the level of operators. In the 4D dictionary, an extrapolated operator takes the form
\be\begin{split}
    & \mathcal{O}^{+}_{\Delta}(z,\bz) \propto\Gamma(\Delta)\lim_{r\rightarrow\infty}\int \d u\, (u-\im\epsilon)^{-\Delta_i}r \Phi(u,r,z,\bz), \\
    & \mathcal{O}^{-}_{\Delta}(z,\bz) \propto\Gamma(\Delta)\lim_{r\rightarrow\infty}\int \d v\, (v+\im\epsilon)^{-\Delta_i}
r \Phi(v,r,z,\bz) ,
\label{eq:extr}
\end{split}
\ee
where the superscript $\pm$ indicate extrapolation to $\scri^\pm$, corresponding to a smearing in $u$ or $v$ respectively.\footnote{\label{ft:iepsuv} Note that the opposite $\im\epsilon$ prescription (i.e. $u+\im\epsilon$ and $v-\im\epsilon$) selects out modes that annihilate the respective in/out vacuum.} As before, we consider the extrapolation to future null infinity without loss of generality. To start, we change to inverted coordinates in the bulk $X^\mu\to\bar{X}^\mu=X^\mu/X^2$, and rewrite everything in terms of inverted operators. First note that
\begin{equation}
    \Phi\left(X^\mu\right)= \Phi\left(\frac{\bar{X}^\mu}{\bar{X}^2}\right)=\bar{X}^2 \,\tilde{\Phi}\left(\bar{X}^\mu\right),
\end{equation}
with
\begin{equation}
    \bar{X}^2=-\frac{1}{u(u+2r )}=-\frac{1}{2u r} +O(r^{-1})\,
\end{equation}
so that in the large-$r$ limit we have 
\begin{equation}
    \bar{X}=-\frac{1}{(u+2r)}\hat{t}-\frac{r}{u(u+2r)}\hat{q}(z,\bar{z})=-\frac{1}{2u} \hat{q}(z,\bar{z})+O(r^{-1}),
\end{equation}
where we have used the parameterization in~\eqref{eq:parY}.
Rewriting the smearing over null infinity in terms of inverted operators and the inverted integration variable $t=u^{-1}$, we thus get
\begin{equation}
     \lim_{r\to\infty}\int_{-\infty}^\infty du\,(u-\im \epsilon) ^{-\Delta} \,r \Phi(u,r,z,\bar{z})
   \sim-2^{\Delta-1}\int_{-\infty}^{\infty}dt \,\frac{\left(\frac{1}{t}-\im\epsilon\right)^{-\Delta}}{t}\tilde{\Phi}\left(-t\,\hat{q}(z,\bar{z})\right).
   \label{eq:intgrlequiv}
\end{equation}
In other words, we see that smearing an operator along null infinity according to \eqref{eq:extr} is equivalent to a Mellin transform of the inverted operator over the light cone with the regularization of the measure encountered in the previous section. Note that the argument on the right hand side is what we expect from the 3D dictionary, namely
\be
\hat{q}(z,\bar{z})=\lim_{\rho\to\infty}\rho^{-1}\hat{Y}(\rho,z,\bar{z}),
\ee
as can bee seen from \eqref{AdS}. Now, we saw  in the previous section that for example when it comes to issues of time ordering, it can be relevant to know from which side of the light cone we are approaching $\hat{q}$. Figure~\ref{fig:inversionscri} gives schematically which regions near $\scri^\pm$ maps to which sides of the light cone under an inversion. Accordingly, we should write \eqref{eq:intgrlequiv} in the following way (suppressing the $\im \epsilon$ in the measure for clarity)
\begin{equation}
\begin{split}
    2^{1-\Delta} \int_{-\infty}^\infty du\,u ^{-\Delta} \lim_{r\to\infty}\,r \Phi(u,r,z,\bar{z})  & \\\
   =  e^{\im  \pi\Delta}&\lim_{\hat{Y}\to Q}\int^\infty_0 dt\,t^{\Delta-1} \tilde{\Phi}(t \hat{Y}_{dS})-\lim_{\hat{Y}\to -Q}\int^\infty_0 dt\,t^{\Delta-1} \tilde{\Phi}(t \hat{Y}_{AdS})
\end{split}\label{eq:equiv}
\end{equation}
where, as discussed above, to match~\eqref{eq:STdict} we have $Q= \rho q$.
Finally, translating this into a statement about extrapolated operators gives
\begin{equation}
   \mathcal{O}^{+}_{\Delta}(Q)_{4D}=2^{\Delta-1}\left(e^{\im  \pi\Delta}\tilde{\mathcal{O}}_{\Delta,\,\text{dS}}(Q)_{3D}-\tilde{\mathcal{O}}_{\Delta,\,\text{AdS}}\left(Q\right)_{3D}\right),
\end{equation}
or, if we perform an inversion on both sides and apply \eqref{eq:shdwop} relating inversions and shadow transformations
\begin{equation}
   \mathbf{S}\left[\mathcal{O}^{+}_{2-\Delta}\right](Q)_{4D}=2^{\Delta-1}\left(e^{\im  \pi\Delta}{\mathcal{O}}_{\Delta,\,\text{dS}}(Q)_{3D}-{\mathcal{O}}_{\Delta,\,\text{AdS}}\left(Q\right)_{3D}\right),
   \label{eq:SO4D3D}
\end{equation}
where we have used the fact that as in~\cite{Strominger:2017zoo} the coordinates on the in and out celestial spheres are antipodally related, i.e. $Q$ and $-Q$ correspond to the same point.

Comparing to our discussion of the wavefunctions in the previous subsection there are a few things worth noting. First, we note that the combination in~\eqref{eq:SO4D3D} matches the one appearing in~\eqref{eq:invSTwf} up to a constant prefactor. Indeed we can also understand the AdS and dS in those combinations via figure~\ref{fig:inversionscri}. Namely, the large $r$ limit in the 4D extrapolate dictionary tells us which side we are approaching the $i^0$ lightcone from. Under the inversion image we get one AdS and one dS contribution, matching~\eqref{eq:invSTwf}. Moreover, we have seen that we can track the 4D $\im \epsilon$-prescription~\eqref{eq:extr} under the inversion. The one inherited from \eqref{eq:extr}, is the $t^{\Delta-1}\to\left(\frac{1}{t}\mp\im\epsilon\right)^{-\Delta}/t$ encountered in the previous section for out- and in-coming operators respectively.

With this matching of the operators under inversions, we can now obtain a relationship between the extrapolation using the 4D dictionary and extrapolation using 3D dictionary. In particular the $S$-matrix elements~\eqref{eq:4Ddict} are correlators of operators~\eqref{eq:extr} related by inversion to the operators in~\eqref{eq:SO4D3D}. The contribution to these correlation functions coming from operators within the first Poincar\'e patch is precisely~\eqref{eq:STdict}.

Now in the previous subsection we saw that  -- with an appropriate contour prescription -- we were able to extract a wavefunction in terms of operators near the lightcone of the origin on a single Poincar\'e patch.  As discussed there, the origin of the contour subtleties was related to the fact that inversions do not preserve the time ordering of the operators.  By contrast the Einstein cylinder time ordering is unaffected. Indeed, one should be wary of the quotient of the Einstein cylinder to one Poincare patch, since we would only expect correlation functions with appropriate `image charges' to be well defined under this quotient. In the next section we will show how to use the cylinder picture to understand how to turn~\eqref{eq:STdict} into an $S$-matrix element. As a nice side effect, we will see how the powers in $r$ we should strip off in the extrapolate dictionary are expected to depend on the bulk scaling dimension. 

\section{Lifting to the Einstein Cylinder}\label{sec:lift}

The goal of this section is to understand how to square the two proposals for extracting $S$-matrix elements from the 4D and 3D extrapolate dictionaries. We will do so by up-lifting our story to the Einstein cylinder. In the process we will also see that this picture elucidates how the bulk scaling dimension affects the extrapolate dictionary~\eqref{eq:4Ddict0}. Let's start by reviewing this uplift.

\paragraph{Correlators on the Einstein cylinder}

To recap, we are studying the celestial holographic dual of a 4D CFT in Minkowski space, focusing on scalar operators $\Phi_{\delta}(X)$ of conformal weight $\delta$.\footnote{Not to be confused with a boundary operator of weight $\Delta$.}  Bulk correlators of such a 4D CFT can be analytically continued from Minkowski space to the Einstein cylinder, essentially because their singularities depend only on the conformal structure. See for example proposition 2 of \cite{Luscher:1974ez} for scalar correlators.  The analytically continued correlators may be viewed as correlators of embedding space operators
\be\label{phideltaZ}
\Phi_\delta(Z) = (Z^{-1}+Z^4)^{-\delta}\,\Phi_\delta(X),
\ee
where one sets $X^\mu = Z^\mu/(Z^{-1}+Z^4)$ on the right. These operators are homogeneous of weight $-\delta$ under projective rescalings of $Z^I$. Because of this, the $\Phi_\delta(Z)$ can be thought of as operators on the projectivization of the 6D null cone $Z^IZ_I=0$ \cite{Rychkov:2016iqz}. They can be pulled back by the covering map \eqref{zm} to operators on the Einstein cylinder. Our various dictionaries are then obtained by placing such operators along various light cones or null infinities on the cylinder.

The main property characterizing $\Phi_\delta(Z)$ is that it transforms as a scalar under conformal transformations of $\bbM=\bbR^{1,3}$. That is, it transforms as the restriction of a Lorentz scalar of $\bbR^{2,4}$ to the quadric:
\be
\Phi_\delta(Z)\mapsto\til\Phi_\delta(W) = \Phi_\delta(Z)
\ee
for any conformal map $Z^I\mapsto W^I$. The Weyl rescaling of the 4D metric \eqref{ds2X} generated by this conformal transformation is $\d W_I\d W^I/(W^{-1}+W^4)^2 = \Omega^2\d Z_I\d Z^I/(Z^{-1}+Z^4)^2$, from which we read off
\be
\Omega = \frac{Z^{-1}+Z^4}{W^{-1}+W^4}
\ee
because $Z^I\mapsto W^I$ is a 6D Lorentz transformation. Hence, \eqref{phideltaZ} ensures that $\Phi_\delta(X)\mapsto\Omega^{-\delta}\Phi_\delta(X)$ as required of a 4D conformal primary. This scalar transformation law continues to hold when $\Phi_\delta(Z)$ is pulled back to the cylinder $\bbE$. Define the cylinder scalar operators
\be
\Phi_\delta(\tau,\vec e) \vcentcolon= \Phi_\delta(Z(\tau,\vec e))\,,
\ee
where we are abbreviating by unit 4-vectors $\vec e$ the 3-sphere positions
\be
\vec e = (\sin\psi\sin\theta\cos\phi,\,\sin\psi\sin\theta\sin\phi,\,\sin\psi\cos\theta,\,\cos\psi)\,.
\ee
The scalar nature of $\Phi_\delta(\tau,\vec e)$ suggests that the extra factor of $r^{\delta}$ should be present from the start in celestial correlators on the cylinder. Let us see how this happens.

Using \eqref{phideltaZ} and the covering map \eqref{zm}, the cylinder operators can be related to the 4D primaries:
\be\label{cyclops}
\Phi_\delta(\tau,\vec e) = (\cos\tau+\cos\psi)^{-\delta}\,\Phi_\delta(X)\,.
\ee
One similarly finds the relation between correlators on the cylinder and correlators on a Minkowski patch:
\be\label{cylcorr}
\big\la\Phi_{\delta_1}(\tau_1,\vec e_1)\cdots\Phi_{\delta_n}(\tau_n,\vec e_n)\big\ra = \prod_{i=1}^n(\cos\tau_i+\cos\psi_i)^{-\delta_i}\;\big\la\Phi_{\delta_1}(X_1)\cdots\Phi_{\delta_n}(X_n)\big\ra\,.
\ee
These extend beyond a single Minkowski patch to the entire Einstein cylinder (at least for unitary CFTs) \cite{Luscher:1974ez}. 

Let's take the large $r$ limit of \eqref{cyclops} in a given Minkowski patch. The extra factor of $r^{\delta}$ needed for taking the large $r$ limit will arise naturally from the rescaling $(\cos\tau+\cos\psi)^{-\delta}$. This is because on the cylinder, null infinities lie precisely along $Z^{-1}+Z^4=\cos\tau+\cos\psi=0$. In the ``fundamental'' Minkowski patch $\tau\in(-\pi,\pi),\psi\in(0,\pi)$, null infinites are the loci $\scri^\pm = \{\psi\pm\tau = \pi\}$. In the large $r$ limit, comparing \eqref{scribondi} with \eqref{zm}, Bondi time on $\scri^+$ can be related to the 3-sphere coordinates $(\psi,\theta,\phi)\in S^3$ by
\be
u = \cot\psi + O\bigg(\frac{1}{r}\bigg)\,.
\ee
That is, we are interpreting the large $r$ limit at fixed $u,z,\bz$ as a limit $\tau\to\pi-\psi$ at fixed $\psi,\theta,\phi$. 

Using the parametrization \eqref{xem} of the interior of the Minkowski patch, one also obtains
\be
r^2 = (X^1)^2+(X^2)^2+(X^3)^2 = \frac{\sin^2\psi}{(\cos\tau+\cos\psi)^2}\,.
\ee
Because $\tau$ is approaching $\pi-\psi$ from below, write $\tau = \pi-\psi-\al$. Then as $\al\to0^+$, we find $\cos\tau+\cos\psi = \al\sin\psi + O(\al^2)$. Since $\psi\in(0,\pi)$, which is the range in which $\sin\psi$ is positive, we see that $\cos\tau+\cos\psi$ also remains positive as $\tau$ approaches $\scri^+$ from below. So we may set
\be
r = \frac{\sin\psi}{\cos\tau+\cos\psi}
\ee
for the purposes of the large $r$ limit.  Eliminating $\cos\tau+\cos\psi$ in terms of $r$, we can finally re-express \eqref{cyclops} as
\be
\Phi_\delta(\tau,\vec e) = (r\csc\psi)^\delta\,\Phi_\delta(X)\,.
\ee
In the large $r$ limit, we may use $u = \cot\psi+O(1/r)$ to replace $\csc\psi$ by $\sqrt{1+u^2} + O(1/r)$. This leads to one of our main results:
\be
\Phi_\delta(\tau,\vec e)\bigr|_{\scri^+} \equiv \Phi_\delta(\pi-\psi,\vec e) = (1+u^2)^{\delta/2}\,\O_\delta(u,z,\bz)\,.
\ee
Here, $\vec e|_{\scri^+}$ is also interpreted as a function of $u,z,\bz$ by inverting $\psi,\theta,\phi$ in terms of Bondi coordinates at $\tau=\pi-\psi$. 

One may now Mellin transform this in $u$ to obtain the corresponding celestial operators
\be
\O_{\delta,\Delta}^\pm(z,\bz) = \int_{-\infty}^\infty\d u\,(u\mp\im\epsilon)^{-\Delta}(1+u^2)^{-\delta/2}\,\Phi_\delta(\tau,\vec e)\bigr|_{\scri^+}\,.
\ee
This establishes that celestial correlators of bulk CFTs may be computed using the uplift of the bulk CFT to the Einstein cylinder by Mellin transforming the position space correlators
\be
\big\la\O_{\delta_1}(u_1,z_1,\bz_1)\cdots\O_{\delta_n}(u_n,z_n,\bz_n)\big\ra = \prod_{i=1}^n(1+u_i^2)^{-\delta_i/2}\;\Big\la\Phi_{\delta_1}(\tau_1,\vec e_1)\bigr|_{\scri^+}\cdots\Phi_{\delta_n}(\tau_n,\vec e_n)\bigr|_{\scri^+}\Big\ra\,.
\ee
This encapsulates the celestial extrapolate dictionary for massless scalars in 4D CFTs in a manner independent of the choice of 4D conformal frame, and it demystifies the relation between the bulk scaling dimension and the conformal factor we need to extract in the extrapolate dictionary. We can now use this uplift to tie together the 3D and 4D extrapolate dictionaries.

\paragraph{Obtaining the S-matrix from $n$-point functions}
Now while the correlators~\eqref{eq:STdict} proposed in~\cite{Sleight:2023ojm} were designed to transform under the correct celestial weights, the authors only claimed to extract the $S$-matrix by cutting the propagators that appear in the correlation function~\eqref{eq:STdict}
\begin{equation}
\langle\mathcal{O}_{\Delta_1}(Q_1)\cdots\mathcal{O}_{\Delta_n}(Q_n) \rangle_{3D, ~\rm{on-shell} }  = \prod_i \lim_{\hat{Y}_i\to Q_i}\int_0^\infty \d t_i\, t_i^{\Delta_i-1}\,\Im_i\langle \Phi(t_1\hat{Y}_1)...\Phi(t_n\hat{Y}_n) \rangle\,.\label{eq:STamp}
\end{equation}
computing the discontinuity 
\be
{\rm Disc}_{P_i^2}\left[\frac{1}{P_i^2-\im\epsilon}\right]=\Im_i\left[\frac{1}{P_i^2-\im\epsilon}\right]=2\pi\im\,\delta(P_i^2)
\ee
before Mellin transforming to the boost basis.\footnote{Note that here we interchanged the Mellin integral and imaginary part in~\eqref{eq:STamp} as compared to~\cite{Sleight:2023ojm} since we are generally interested in complex weights.}

Now we know from the work of~\cite{He:2020ifr} that the usual procedure of cutting the on-shell propagators in the LSZ formulation indeed maps to the 4D extrapolate dictionary, which gives the S-matrix element $A_n$ in terms of boundary operators at null infinity. Mellin transforming the momentum eigenstate result in~\cite{He:2020ifr} to the boost basis, one concludes the the LSZ prescription picks out~\cite{Pasterski:2021dqe}
\begin{equation}
    A_n\sim \langle T\{\mathcal{O}^1_{\Delta_1}\dots \mathcal{O}^n_{\Delta_n} \} \rangle_{4D},
\end{equation}
where each operator is the difference of a future and past extrapolated operator
\begin{equation}\label{eq:doubleinout}
\badat{3}
    \mathcal{O}^{+,i}_{\Delta_i, 4D} =   \mathcal{O}^{+,i}_{\Delta_i,4D}|_{\scri^+} - \mathcal{O}^{+,i}_{\Delta_i,4D}|_{\scri^-},\\
       \mathcal{O}^{-,i}_{\Delta_i, 4D} =   \mathcal{O}^{-,i}_{\Delta_i,4D}|_{\scri^-} - \mathcal{O}^{-,i}_{\Delta_i,4D}|_{\scri^+}.
       \eadat
\end{equation}
More explicitly
\begin{equation}\label{eq:outdouble}
    \mathcal{O}^{+,i}_{\Delta_i, 4D}   \propto\Gamma(\Delta)\lim_{r\rightarrow\infty}r \left[\int du (u-\im\epsilon)^{-\Delta_i}\Phi(u,r,z,\bz)-\int dv (v-\im\epsilon)^{-\Delta_i}
 \Phi(v,r,z,\bz)\right].
\end{equation}
Here we've introduced an additional label in~\eqref{eq:doubleinout} to distinguish the $\im\epsilon$ prescription from the (future/past null infinity) location of the operators, which are often conflated by virtue of the comment in footnote~\ref{ft:iepsuv}. Namely, the second term in~\eqref{eq:outdouble} is expected to annihilate the $in$ vacuum. Thus the $S$-matrix element can be computed as a correlation function of just the first operator appearing on the right hand side of each line in~\eqref{eq:doubleinout}, namely the 4D dictionary we discussed above~\eqref{eq:4Ddict} with operators~\eqref{eq:extr}.

We can also lift this story to the Einstein cylinder.  The $S^3$ slice of the cylinder through $i^0$ is also a Cauchy slice of the conformally compactified Minkowski space embedding in the corresponding Poincar\'e patch. As such one can prepare the vacuum state by either evolving to the infinite past on the cylinder (i.e. no operators inserted before the Poincar\'e patch) or with a path integral over a Euclidean cap as in~\cite{Chen:2023tvj}. Then the scattering $|in\rangle$ state can be prepared by acting with the operators at the locus corresponding to past null infinity of that Poincar\'e patch.  An analogous construction prepares the out state from the out vacuum by acting with operators at the locus corresponding to future null infinity.\footnote{See appendix~\ref{app:conjugation} for discussion of the RSW $\langle out|$ state~\cite{Cotler:2023qwh}. }

\begin{figure}
\centering
\hspace{-10cm}

  \adjustbox{max width=.45\textwidth}{
        \begin{tikzpicture}[scale=11]
    \hspace{1cm}
    \draw[fill=cyan, fill opacity=.2]  (.5,.5) -- (1,0) -- (.5,-.5) -- cycle;
    \draw[gray]  (.5,.5) to [bend left=45]  (.5,-.5) ;
    \draw[fill=red]  (.705,.06) circle (0.02cm) ;
    \draw[-latex,line width= 3]   (.705+.02,.06+.02) --  (.705+.02+0.1,.06+.02+0.1);

     \begin{scope}[decoration={markings,mark=at position 0.5 with {\arrow[scale = 3]{latex}}}]
    \draw[red,postaction={decorate}] (1,.0) -- (0.5,0.5) ;
    \end{scope}
    \node[] at (.8,-.4) {\Large \textbf{(A)}};
\end{tikzpicture}
\qquad
 \begin{tikzpicture}[scale=11]
    \hspace{1cm}
    \clip (.495,-.5) rectangle (1,.5) ;
    \draw[fill = magenta, fill opacity = .2
]  (.5,.5) -- (1,0) -- (.5,-.5) -- cycle;
    \draw[black] (.5+0.25,-.25) -- (.5,.0) -- (.5+0.25,.25); 
     \begin{scope}[decoration={markings,mark=at position 0.75 with {\arrow[scale = 2]{latex}},pre=curveto,post=curveto},decorate]
    \draw[blue,postaction={decorate}] (.5-.25,.25) to [bend left = 45] (.5+.25,.25);
    \draw[blue,postaction={decorate}] (.5-.25,.25) to [bend left = 45/2] (.5+.25,.25);
     \draw[blue,postaction={decorate}] (.5-.25,.25) to [bend left = 0] (.5+.25,.25);
    \draw[blue,postaction={decorate}] (.5-.25,.25) to [bend left = -45/2] (.5+.25,.25);
    \draw[blue,postaction={decorate}] (.5-.25,.25) to [bend left = -45] (.5+.25,.25);
    \end{scope}

     \begin{scope}[decoration={markings,mark=at position 0.55 with {\arrow[scale = 2]{latex}},pre=curveto,post=curveto},decorate]
    \draw[blue,postaction={decorate}] (.75,.25) to [bend left = 45] (.75,-.25);
    \draw[blue,postaction={decorate}] (.75,.25) to [bend left = 45/2] (.75,-.25);
     \draw[blue,postaction={decorate}] (.75,.25) to [bend left = 0] (.75,-.25);
    \draw[blue,postaction={decorate}] (.75,.25) to [bend left = -45/2] (.75,-.25);
    \draw[blue,postaction={decorate}] (.75,.25) to [bend left = -45] (.75,-.25);
    \end{scope}

    \begin{scope}[decoration={markings,mark=at position 0.4 with {\arrow[scale = 3]{latex}}}]
    \draw[red,postaction={decorate}] (.5,.0) -- (.75,.25) ;
    \end{scope}

    \begin{scope}[decoration={markings,mark=at position 0.2 with {\arrow[scale = 3]{latex}}}]
    \draw[red,postaction={decorate}] (.5,.0) -- (.75,-.25) ;
    \end{scope}

    \draw[fill=red]  (.6,-.1) circle (0.02cm) ;

    \node[] at (.8,-.4) {\Large \textbf{(B)}};
  
\end{tikzpicture}
\raisebox{.0cm}{
   \begin{tikzpicture}[scale=3.2]
\hspace{1.5cm}

  \draw (0,-.64) ellipse (.73 and .25);
    \draw (0,2.1) ellipse (.73 and .25);
     \draw[rotate=-45,fill=cyan, fill opacity=.1] (0,0) ellipse (1 and .18);
     \begin{scope}
  \clip (-.73,-1) rectangle (.73,3);
    \clip (3.4,-2) circle (5);
     \clip (0,0) circle (2.24);
     
    \clip  (-.65,0.5)--(-.73,0.6)--(-.73,0.73) --(-.63,1) -- (-.7,1.5) -- (1,2.3) -- (1,-.73) --  (.3,-.7) -- cycle;
  \fill[cyan,fill opacity=.2] (3.4,3.4) circle (5);
\end{scope}

 \draw[red,rotate=45,fill=white, fill opacity=.5] (1,1) ellipse (1 and .18);
        \draw[rotate=-45] (-1,1) ellipse (1 and .18);
     \draw[rotate=45] (0,0) ellipse (1 and .18);
     \draw[] (-.73,-.64) -- (-.73,2.1);
     \draw[] (.73,-.64) -- (.73,2.1);

     \begin{scope}
  \clip (.73,-1) rectangle (-.73,3);
    \clip (-3.4,-2) circle (5);
     \clip (0,0) circle (2.24);
     
    \clip  (.65,0.5)--(.73,0.6)--(.73,0.73) --(.63,1) -- (.7,1.5) -- (-1,2.3) -- (-1,-.73) --  (-.3,-.7) -- cycle;
  \fill[magenta,fill opacity=.2] (-3.4,3.4) circle (5);
   
\end{scope}
\node[] at (1.1,-.64) {\Large \textbf{(C)}};
    \draw[fill=red]  (-.2,1.) circle (0.073 cm) ;
\end{tikzpicture}
}
}
\caption{\small A sketch of the two extrapolate dictionaries. \textbf{(A)} the 4D extrapolation procedure: an operator (red dot) is pushed to null infinity by taking $r\to\infty$ with $u$ fixed, before smearing along the null direction (red line) according to \eqref{eq:4Ddict}. \textbf{(B)} the corresponding operation in the 3D dictionary: an operator is smeared along a ray (red line) traversing a family of hyperbolic slices (blue curves) according to \eqref{eq:STdict} before taking the boundary limit in the 3D slice geometry $\hat{Y}\to Q$. In this limit, the operator gets smeared along the light cone of the origin. \textbf{(C)} both procedures lifted to the cylinder on patches that are related by an inversion. The smearing along null infinity and the Mellin transform along the light cone become the same operation on the cylinder.}\label{fig:dicts}
\end{figure}
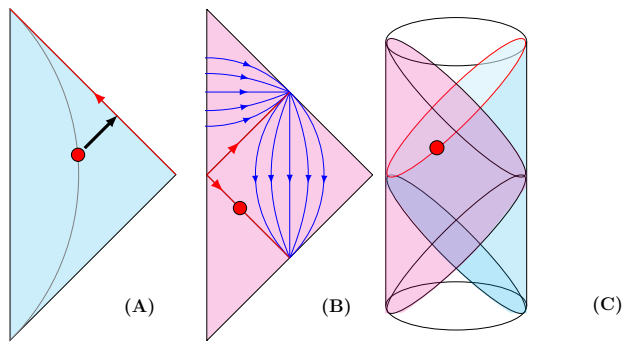

 One of the appeals of these extrapolate dictionaries is that they give us a way to construct an $S$-matrix like object when the bulk itself is a CFT. Part of the reason why one would usually not talk about an $S$-matrix in the context of CFT is that the choice of Poincare\'e patch breaks some of the conformal symmetries.\footnote{Efforts to define a massless $S$-matrix are also plagued with subtleties related to IR divergences in gauge theory and gravity. Part of the motivation for the detector operator literature is to use bulk CFTs to identify good IR finite scattering observables~\cite{Caron-Huot:2022eqs}.} Namely only the Poincar\'e subgroup preserves the Poinca\'e patch. However, this is precisely what we can use to relate the 3D and 4D extrapolate dictionaries. Inversion symmetry of the vacuum correlators of the 4D CFT on the cylinder 
\be
\langle I {\cal O}_{\Delta_1} I^{-1}...I {\cal O}_{\Delta_n} I^{-1}\rangle 
=\langle  {\cal O}_{\Delta_1} ... {\cal O}_{\Delta_n} \rangle 
\ee
is precisely what equates the 4D extrapolate dictionary, \eqref{eq:4Ddict} and \eqref{eq:SO4D3D}, involving correlators of operators on the lightcone of $i^0$, the 3D extrapolate dictionary involving correlators of operators smeared along the lightcone of the origin (see figure~\ref{fig:dicts}). Except, on the cylinder we see that~\eqref{eq:STdict} is only the contribution from the first patch. 
Now from our discussion in section~\eqref{sec:sheqinv} we would conclude that 
\be\label{sheq}
\langle ( {\rm sh}\circ {\cal O})_{\Delta_1} ...( {\rm sh}\circ {\cal O})_{\Delta_n}\rangle=\langle  {\cal O}_{\Delta_1} ... {\cal O}_{\Delta_n} \rangle.
\ee
While the shadow wavefunctions are different from the standard Mellin transform basis, unlike our wavefunction-from-propagator discussion in section~\ref{sec:twoextrap}, we are shadowing every operator appearing in the correlator. The fact that the shadow squares to one, takes us back to a contact term 2-pt function when both operators are shadow transformed, so we see that~\eqref{sheq} indeed holds for the free theory. This argument would also seem to hold for any 4D CFT. Meanwhile, amplitudes in the shadow basis for non-conformal theories have been studied here \cite{Chang:2022jut}.

Since we know the 4D $S$-matrix is equal to the correlation function with the operators extended to the next Poincar\'e patches, if we trust the construction of the $S$-matrix in~\cite{Sleight:2023ojm} we have an interesting relation between computing discontinuities of the correlation functions in the single Poincar\'e patch and the correlators uplifted to the Einstein cylinder that extend beyond the first Poincar\'e patch. This has an intriguingly similar flavor to the discussion of Regge correlators, detector operators and OTOCs in~\cite{Caron-Huot:2022lff}.

However, we also want to reconcile this with our discussion of the wavefunctions for the 3D extrapolate dictionary on the single Poincar\'e patch in section~\ref{sec:wfprop}. While one can follow a similar route as in~\eqref{sec:wfprop} with one point in a fixed Poincare\'e patch (say, the blue on in figure~\ref{fig:dicts}) and the other smeared along the operators insertions at either the (extended) lightcone of the origin, here we will discuss a simpler argument related to the construction in~\cite{Chen:2023tvj}. To be precise, rather than needing the full uplift to the cylinder, the question is how to relate the wavefunctions in the pink and blue patches to each other in figure~\ref{fig:dicts}c. There the authors of~\cite{Chen:2023tvj} use the fact that on the coordinates on the overlap (purple shaded region, which contains a Cauchy slice for both patches) are related by an inversion and the corresponding wavefunctions on this region are related by a shadow to then analytically continue to the remainder of the two antipodal Poincar\'e patches. By noting that the $\mathbb{Z}_2$ quotient isn't necessary for this region because points stay within the original Poincar\'e patch, we see that the contour choice in section~\ref{sec:wfprop} indeed produces the inverted wavefunctions (with any sources outside of the intersection of the interiors of these two patches), as necessary for this cylinder-based argument.  It will be interesting to further examine how the contour choice in section~\ref{sec:wfprop} more generally amounts to computing the correlation functions with operators extending beyond the first Poincar\'e patch.

In sum, we've see here that we can use the uplift picture to both understand the power of $r$ in the 4D extrapolate dictionary and relate different extrapolate dictionaries. While here we have focused on two convenient conformal frames related by a 4D inversion --  which has the feature of preserving the Lorentz subgroup and, hence, boost eigenstate property of the `celestial operators' -- this construction naturally extends to other `celestial'-like dictionaries related by 4D conformal transformations to these, and we explore this further via another example that readily connects to the conformal collider literature in appendix~\ref{app:gendict}. Namely if we are interested in 4D CFTs or perturbative computations around the free bulk theory, we can start with ordinary correlators of operators placed along a light-sheet in the bulk, and conformally transform them to celestial correlators without the need to ever take an explicit large $r$ limit. Let's now turn to a discussion of our results in this paper and natural future directions and connections.

\section{Discussion}\label{sec:discussion}

The role of conformal inversions is being increasingly appreciated in celestial holography. Besides the hints from the explicit form of the wavefunctions in~\cite{Pasterski:2017kqt}, more recently, inversions played a key role in the top-down celestial holographic duality of \cite{Costello:2022jpg,Costello:2023hmi}, relating observables on Burns space to observables on $\CP^2$ \cite{Hawking:1979pi}. They were also used in the study of entanglement entropy in the celestial CFT dual to Maxwell's theory \cite{Chen:2023tvj}.

Motivated by these and related facts, in this work we studied the effects of bulk conformal transformations on state and operator dictionaries in celestial holography. We mostly focused on conformal inversions and used the massless free scalar as our running example. The latter is automatically conformally coupled in flat space, so conformal transformations map solutions of the flat space Klein-Gordon equation to new solutions. In particular, we showed how conformal inversions map wavefunctions of boost eigenstates of weight $\Delta$ to shadow transforms of boost eigenstates of weight $2-\Delta$. As our main result, in the massless case, this allowed for an identification of the Sleight-Taronna celestial correlators \cite{Sleight:2023ojm} with shadow transforms of standard celestial correlators.

A crucial feature of our story has been the fact that conformal transformations can be used to map null infinity to null surfaces in the interior of the 4D Minkowski bulk. This becomes clearest in the compactification of Minkowski space on the Einstein cylinder, where null infinity can be treated on the same footing as any other points in the bulk. The usual extrapolate dictionaries of AdS holography \cite{Harlow:2011ke} equate correlators of bulk operators extrapolated to the boundary of spacetime with correlators of a boundary dual, and this is widely believed to continue to hold in celestial holography. In the light of such an extrapolate dictionary, bringing null infinity to a finite light cone or light-sheet within the bulk has the curious feature of mapping the holographic plate to the interior of the bulk, at least when we are computing things in perturbation theory or for a 4D CFT. 

Let us close by compiling some further observations and directions for future work.

\paragraph{More fun with Detectors} We will start by expanding on some of the things we learned and speculated on in the last section. First, we saw that the uplift to the Einstein cylinder provided a unified picture of the various holographic dictionaries related by conformal transformations. This could be applied either to construct an $S$-matrix like object for a 4D CFT or to describe perturbative amplitudes more generally, for the case where the 4D CFT is the free theory. On the one hand these manipulations are convenient for understanding celestial OPEs and adopting technology and results already developed in the conformal collider literature~\cite{Hofman:2008ar,Cordova:2018ygx} to the celestial hologram~\cite{Hu:2022txx,Hu:2023geb}. Meanwhile we see how the extrapolate dictionary gets modified when the bulk scaling dimension runs, which is relevant to the weakly coupled detector story~\cite{Caron-Huot:2022eqs} and provides some insight into how this modification to the expected radiative falloffs would affect what we mean by the scattering data.

While both of these programs look at boost eigenstates, while the celestial literature focuses on in and out states with a finite number of particles (and usually starts from $S$-matrix elements where none of them are collinear), the detector/collider literature is often interested in inclusive quantities. The fact that our cylinder set up can let us natural phrase things in terms of a path integral~\cite{Arefeva:1974jv,Kim:2023qbl} for different boundary conditions of the fields at the boundary of a Poincar\'e patch, makes this natural to do. Indeed continuity on the cylinder makes the final (with respect to cylinder time) conditions on the first Poincar\'e patch match the initial conditions on the next Poincar\'e patch, making it tempting to consider inclusive quantities in terms of an evolution across two Poincar\'e patches (i.e. spaced by $2\pi$ in cylinder time). Indeed our discussion of how the 3D dictionary relates to the 4D construction of the $S$-matrix seemed to also hint at this in connection to certain simple time contours, e.g. in the camera story in~\cite{Caron-Huot:2022lff} as opposed to the more complicated ones related to crossing in~\cite{Caron-Huot:2023vxl}. It would be interesting to explore this further and understand what $S^\dagger S=1$ looks like from this cylinder perspective or from embedding space.

\paragraph{Matching different flat space limits of AdS$_4$/CFT$_3$.}
Now while the discussions above could apply to either a general 4D CFT or perturbation around the free theory, let us restrict to the latter case and compare our 3D and 4D extrapolate dictionary expressions for the $S$-matrix to what one might get from the flat limit of an AdS$_4$ bulk. There are two proposals that at first glance seem a bit different, but seem to be naturally connected  -- upon taking the the flat limit -- in this set up. First, the construction in~\cite{Hijano:2019qmi} involves Fourier transforming operators on the conformal boundary in the time coordinate near the window probed by the bulk point
\be
S(p_i)=\lim\limits_{l\rightarrow\infty}l^{\frac{D-3}{2}}\left[\prod_{i}C(p_i)\int dt_ie^{\pm i\omega_it_i}\right]\langle 0|{\cal O}(\tau_1,\Omega_1)...{\cal O}(\tau_n,\Omega_n)|0\rangle,
\ee
where $D=4$ is the bulk dimension, the bulk scaling dimensions are $\delta=\mathcal{O}(l^0)$ in the flat limit, and the operators are smeared near the bulk points 
\be
\tau_i=\pm\frac{\pi}{2}+\frac{t_i}{l},\ee
and placed at directions corresponding to the momentum direction. This one is closest to the 4D extrapolate dictionary. Indeed, as was discussed in the recent work~\cite{deGioia:2023cbd}, and the upcoming~\cite{kp, cpt}, the flat limit of bulk takes a corresponding Carrollian limit of the boundary in which case the regions near the future and past lightcone of the bulk point map to future and past null infinity~\cite{Hijano:2019qmi}, and the Fourier transform in $t$ becomes a Fourier transform in $u$ and $v$ that prepares momentum eigenstates. The celestial basis is just the composition of this with the usual Mellin transform~\cite{deBoer:2003vf,Pasterski:2017kqt}.

Meanwhile in~\cite{Komatsu:2020sag} (see also~\cite{Penedones:2010ue,Gary:2009ae,Fitzpatrick:2010zm}) the momentum space $S$-matrix is extracted from a limit of position space correlators
\be
\lim\limits_{R\rightarrow\infty}\langle {\cal O}_1(x_1){\cal O}_2(x_2)...\rangle=\langle {\vec{p}_1\vec{p}_2...|...\vec{p}_n}\rangle,
\ee
where the $x_i$ are related to a certain analytic continuation of the boundary points. Albeit for massive particles, their construction in terms of bulk to boundary propagators in AdS is more directly related to the 3D dictionary construction in~\cite{Sleight:2023ojm}. While the Sleight-Tarrona proposal is inspired by the hyperbolic foliation of~\cite{deBoer:2003vf}, their final expressions~\eqref{eq:STdict} and~\eqref{eq:STamp} are derived using a Mellin transform of the 4D position space propagator. of in the manner that their construction is also starting from the 4D propagator. Indeed, if we inverse Mellin transform both sides of~\eqref{eq:STdict} and~\eqref{eq:STamp} -- i.e. we undo their change of basis to the boost-eigenstate primary operators --  then we seem to have a position space correlator whose discontinuity is related to the momentum space amplitude. Notably the inversion is exchanging the measure used for the momentum space $\omega$ of a massless particle vs the position space $u$ coordinate for an operator at null infinity. The perspective gives more motivation to understand the role of the discontinuity /  contribution from outside the first Poincar\'e patch.

All in all, it would be fun to flesh out the connections between these and the Celestial construction in more detail with this relation between different CCFT dictionaries in mind.

\paragraph{Celestial holography in the self-dual sector.} And now for a final application a bit beyond what was under discussion here. One important case where the bulk theory under study happens to be a 4D CFT (albeit a non-unitary one) is self-dual Yang-Mills (SDYM) coupled to a fourth-order axion. This was studied in \cite{Costello:2022wso,Costello:2022upu,Bu:2022dis} and describes celestial holography as a Chern-Simons/WZW-type correspondence. Although this setup is generally studied in Euclidean or split signature flat space, we can still introduce celestial extrapolate dictionaries in the uplift of the SDYM + axion theory to twistor space. The latter has a boundary structure, much like the boundary of spacetime, and one can construct a chiral algebra of bulk operators in the limit as they approach the boundary using the formalism of \cite{Costello:2020ndc}.

To be precise, twistor space is $\mathbb{PT}=\bbR^4\times\CP^1$. Using the Penrose transform, one can find a 6D theory on $\mathbb{PT}$ whose compactification along $\CP^1$ produces SDYM + axion on $\bbR^4$ \cite{Costello:2021bah}. On the other hand, if one decomposes $\bbR^4-0=\bbR_+\times S^3$ and KK reduces the 6D theory along $S^3$, one finds a 3d theory on $\bbR_+\times\CP^1$. This has a natural radial coordinate $r\in\bbR_+$, and the celestial OPEs of the SDYM + axion system are found as the boundary chiral algebra of this 3d theory that lives on the $\CP^1$ at $r\to\infty$. For a CFT, we can use a conformal inversion to exchange $r=\infty$ with $r=0$. This is expected to relate the chiral algebra on the $\CP^1$ at the $r\to0$ boundary to the chiral algebra on the $\CP^1$ at the $r\to\infty$ boundary.

In fact, there's a second conjectural relation between the chiral algebras at $r=0$ and $r=\infty$. They are expected to be `Koszul dual' to each other; see \cite{Costello:2022wso,Paquette:2021cij,Costello:2020jbh} for reviews of this fact. Physically, this means that the chiral algebra at $r=\infty$ is isomorphic to the most general chiral algebra that can be gauge invariantly coupled to the chiral algebra at $r=0$. It is plausible that conformal inversions could provide a concrete means of proving this isomorphism. This will require carefully mapping operators in the extrapolate dictionary at $r=\infty$ to operators living at $r=0$. Understanding such a mechanism for Koszul duality will no doubt be an important step in generalizing such connections between celestial holography and twistor theory.

\section*{Acknowledgements}

We are grateful to Simon Caron-Huot, Jordan Cotler, Yangrui Hu, Justin Kulp, Prahar Mitra, and Rob Myers for useful discussions. The research of SP is supported by the Celestial Holography Initiative at the Perimeter Institute for Theoretical Physics and by the Simons Collaboration on Celestial Holography. AS is supported by a Black Hole Initiative fellowship, funded by the Gordon and Betty Moore Foundation and the John Templeton Foundation. SP and EJ's research at the Perimeter Institute is supported by the Government of Canada through the Department of Innovation, Science and Industry Canada and by the Province of Ontario through the Ministry of Colleges and Universities.

\appendix

\section{Discrete symmetries, dual bases, and all that}\label{app:conjugation}

In this appendix we explore how the computations in section~\ref{sec:geocon} can provide insight into the definition of out states and inner products  
\cite{Pasterski:2019ceq}, relevant to the celestial state-operator correspondence~\cite{Crawley:2021ivb,Cotler:2023qwh} and defining a radially quantized celestial CFT.

In~\cite{Crawley:2021ivb} the authors define an inner product involving a shadow transform on the out states which they claim equips the CCFT with a BPZ like inner product in particular with regards to the  conjugation of the Lorentz generators $L^{c}_{n}=L_{-n}$ unlike what we expect from unitarity of the 4D Lorentz generators Namely 
\begin{equation}
\label{embedding_generators}\scalemath{0.95}{
    \badat{3}
 L_0&=\frac{1}{2}(J_3-i K_3) \, ,~~ & L_{-1}&=\frac{1}{2}(-J_1+i J_2+i K_1+K_2)
  \, ,~~
    & L_1&=\frac{1}{2} (J_1+i
   J_2-i K_1+K_2)
    \, ,~~
   \\
 \bar L_0&=\frac{1}{2} (-J_3-i K_3)
  \, ,~~&
   \bar L_{-1}&=\frac{1}{2} (J_1+i J_2+i K_1-K_2)
    \, ,~~
   & \bar L_1 &=\frac{1}{2} (-J_1+i J_2-i K_1-K_2)
    \, ,~~
   \\
\eadat}
\end{equation}
behave as follows 
\be
L_{i}^\dagger=-\bar{L}_i,~~~\bar{L}_{i}^\dagger=-L_i,
\ee
under the 4D Hermitian conjugation. In~\cite{Cotler:2023qwh} they revisit this inner product, defining a so called RSW inner product in their exploration of an integer basis. This involves a reflection $X^3\mapsto-X^3$, shadow, and Weyl transformation on the out state.

Our understanding of the relation between shadow transformations and inversions can help us unpack these statements a bit. We saw above that on the single particle states $I=\rm{sh}\circ W$, so the shadow and Weyl reflection in the RSW inner product can be replaced with an inversion.  However the fact that $I$ doesn't change $M_{\mu\nu}$ under conjugation~\eqref{eq:IM} means that modifying the Hermitian conjugation condition with an inversion alone will not match that expected for a radially quantized 2D CFT. 
Indeed, while the 2D inversion normally used in discussion of BPZ inner product flips north and south poles,
the 4D inversion does not, as we can see from~\eqref{eq:l0same}-\eqref{eq:l1same} or the fact that the inverted wavefunctions have the same reference direction.

The $X^3$ reflection of the RSW inner product serves this role. Under a parity transformation the rotation and boost generators transform as\footnote{Note that we have only been talking about a real scalar field, but composing with an additional charge conjugation would also give the desired~\eqref{eq:new_conj}.}
\be
PJ^i P^{-1}=J^i~~~PK^i P^{-1}=-K^i
\ee
while the reflection in~\cite{Cotler:2023qwh} is related by a $\pi$ rotation in the $1,2$-plane
\be
RJ^{A} R^{-1}=-J^A,~~RJ^{3} R^{-1}=J^3,~~PK^A P^{-1}=K^A~~PK^3 P^{-1}=-K^3,~~A=\{1,2\}
\ee
so that under the conjugation $O^c=R O^\dagger R$
\be\label{eq:new_conj}
L_i^c =L_{-i},~~~\bar L_i^c =\bar L_{-i}.
\ee
We see that the reflection in the RSW inner product of~\cite{Cotler:2023qwh} gives the correct conjugation properties to treat matrix elements of the holomorphic and anti-holomorphic subalgebras separately, while the shadow makes the two point functions non-distributional.

Going back to the Einstein cylinder picture in section~\ref{sec:geocon}, we see from~\eqref{eq:ishw} and~\eqref{eq:xProt}  the RSW conjugation (or, perhaps aptly rename-able IR conjugation) $\langle\langle \psi|=(IR|\psi\rangle)^\dagger$ corresponds to the following rotation 
\be
IR: ~~\tau\mapsto\tau+\pi,~~\phi\mapsto \phi+\pi
\ee
up to the quotient~\eqref{identified} or, equivalently, the simultaneous reflection $Z^4\mapsto-Z^4$, $Z^3\mapsto -Z^3$ -- implemented with a $\pi$ rotation in the 3,4-plane -- in embedding space. Curiously this matches the time separation that appear in the bulk point singularity story~\cite{Gary:2009ae}, though in that context there is an AdS$_5$ bulk.

\section{Massless scalars in generic dimensions}\label{app:gendscalar}
Here we consider the example of the massless scalar in arbitrary dimensions, since it serves as a nice cross check to see how the extrapolate inversion shadow story generalizes beyond $D=4$. 

In a $D$-dimensional bulk with Lorentzian metric $g$ of mostly plus signature, the conformally covariant version of the massless Klein-Gordon operator is
\be
Y_g = \square - \frac{(D-2)}{(D-1)}\,\frac{R}{4}
\ee
where $R$ again is the Ricci scalar. As before, in flat space with Minkowski metric $\eta$, this reduces to $Y_\eta = \square$. Under a conformal transformation $g\mapsto\Omega^2 g$, it obeys
\be\label{eq:Yop}
Y_{\Omega^2g}(\Omega^{\frac{2-D}{2}}\varphi) = \Omega^{-\frac{D+2}{2}}\,Y_g\varphi\,.
\ee
For $D=4$, this gives the special case \eqref{conflaw}. Now the bulk scaling dimension $\delta=\frac{D-2}{2}$.

From~\eqref{eq:Yop} we see that if $\varphi(X)$ solves the $D$-dimensional massless Klein-Gordon equation, its inversion image
\be
\til\vphi(X) = \frac{1}{|X|^{D-2}}\,\varphi\bigg(\frac{X}{|X|^2}\bigg)
\ee
provides us with an equally good solution. Under inversions, the weight $D-2-\Delta$ boost eigenstate
\be
\vphi_{2-\Delta}(X) = \frac{1}{(\veps q\cdot X)^{D-2-\Delta}}
\ee
maps to the shadow transform of the weight $\Delta$ boost eigenstate:
\be
\til\vphi_{D-2-\Delta}(X) = \frac{|X|^{D-2-2\Delta}}{(\veps q\cdot X)^{D-2-\Delta}}\propto\mathbf{S}[\vphi_\Delta(X)]\,,
\ee
which is indeed consistent with equation (2.23) of~\cite{Pasterski:2017kqt} where there $d=D-2$. 

This generalizes our $D=4$ derivation in section~\ref{sec:sheqinv}. One can also apply the extrapolate dictionary to study the wavefunctions and correlators as in section~\ref{sec:twoextrap}. 
The power of $r$ one needs to strip off is now $\lim\limits_{r\rightarrow\infty}r^{\frac{d}{2}}\phi(u,r,x^A)$, consistent with our discussion in section~\ref{sec:lift} where we saw how it should be related to the bulk scaling dimension.

\section{Building new dictionaries}\label{app:gendict}

Every bulk conformal transformation that does not stabilize null infinity can be used to map the 4D extrapolate dictionary to a new extrapolate dictionary. Suppose $\scri^+$ is mapped to a null surface $\cN$. 
Then the new dictionary will simply equate CCFT correlators with bulk CFT correlators of operators placed along $\cN$. What was special about the original conformal frame and the one related by inversion is that they preserve the Lorentz subgroup and as such prepare boost eigenstates in both cases.\footnote{And hence more properly called CCFT perhaps, though generally we are dimensionally reducing the null hypersurface to its cross section, which we can still call a celestial sphere.} However, other choices of conformal frame are more convenient for understanding the OPEs of celestial operators. In this appendix we will look at a particular example where we map null infinity to a null hyperplane within a single Poincar\'e patch (exchanging the origin and the $u=z=\bz=0$ point on the celestial sphere)  studied by Hoffman and Maldacena~\cite{Hofman:2008ar} and used for examining the appearance of BMS symmetries in generic unitary 4D CFTs by Cordova and Shao~\cite{Cordova:2018ygx} (see also~\cite{Hu:2022txx,Hu:2023geb}).

\subsubsection*{Scattering as correlators on a light-sheet}

Let us study the case of the Hofman-Maldacena map. This is the conformal transformation $X^\mu\mapsto Y^\mu$ given by the map
\be\label{eq:HM}
Y^+ = -\frac{1}{X^+}\,,\quad Y^- = - \frac{|X|^2}{X^+}\,,\quad Y^1 = \frac{X^1}{X^+}\,,\quad Y^2 = \frac{X^2}{X^+}
\ee
where $X^\pm = X^0\pm X^3$ and similarly $Y^\pm=Y^0\pm Y^3$. It finds applications in the construction of detector operators for conformal collider physics \cite{Hofman:2008ar} as well as their celestial generalizations \cite{Hu:2022txx}. 

In embedding space, this conformal map lifts to the Lorentz transformation $Z^I\mapsto W^I$ given by (up to the usual $\bbZ_2$ ambiguity)
\be
W^I = \left(Z^0,-Z^{-1},Z^1,Z^2,-Z^4,Z^3\right)\,.
\ee
It may be checked that this satisfies $Y^\mu = W^\mu/(W^{-1}+W^4)$. In particular, it maps null infinity $Z^{-1}+Z^4=0$ to the locus $W^0+W^3=0$, which is just the light-sheet $Y^+ = 0$. Rather than needing to take a large-$r$ limit to the conformal boundary of a Poincar\'e patch, this is a surface at finite distance: the origin $Y^\mu=0$ lies on it. In the Bondi parametrization \eqref{scribondi} of null infinity (uplifted to $S^1\times S^3$ with say positive sign), one finds that a point $(u,z,
\bz)$ on $\scri^+$ maps to the point
\be\label{Ybondi}
Y^+ = 0\,,\quad Y^- = u(1+|z|^2)\,,\quad Y^1+\im Y^2 = z
\ee
on the light-sheet.

This transforms the 4D Minkowski metric \eqref{ds2X} into
\be
\frac{\eta_{IJ}\d W^I\d W^J}{(W^{-1}+W^4)^2} = \left(\frac{Z^{-1}+Z^4}{Z^0+Z^3}\right)^2\frac{\eta_{IJ}\d Z^I\d Z^J}{(Z^{-1}+Z^4)^2}\,.
\ee
So it is a conformal transformation $\eta_{\mu\nu}\mapsto\Omega^2\eta_{\mu\nu}$ with conformal factor
\be
\Omega = \frac{Z^{-1}+Z^4}{Z^0+Z^3} = \frac{1}{X^+}\,,
\ee
having substituted for $Z^I$ in terms of $X^\mu$ from \eqref{embed}.

Clearly this is different from the conformal inversions we studied above, but we can nevertheless use it to generate a new celestial holographic dictionary. The scalar of weight $\delta$ transforms as a conformal primary:
\be
\Phi_\delta(X)\mapsto\Omega^{-\delta}\,\Phi_\delta(X) =  (X^+)^\delta\,\Phi_\delta(X)\,.
\ee
For example, for the massless free scalar $\Phi(X)$ discussed in the previous section, the conformal weight is $\delta=1$ (in four dimensions) and the associated transformation law is $\Phi\mapsto X^+\Phi$.

To construct a new dictionary, define the new scalar operators
\be\label{PhiY}
\til\Phi_\delta(Y) = (X^+)^\delta\,\Phi_\delta(X),
\ee
where the right hand side is understood as a function of $Y^\mu$ by using the inverse of the Hofman-Maldacena map,
\be
X^+ = -\frac{1}{Y^+}\,,\quad X^- = -\frac{|Y|^2}{Y^+}\,,\quad X^1 = -\frac{Y^1}{Y^+}\,,\quad X^2 = -\frac{Y^2}{Y^+}\,.
\ee
Here, $|Y|^2=\eta_{\mu\nu}Y^\mu Y^\nu = -Y^+Y^-+(Y^1)^2+(Y^2)^2$ as usual. 

We would like to study correlators with the new operators $\til\Phi_\delta(Y)$ placed along the light-sheet $Y^+=0$. Since $\scri^+$ gets mapped to this light-sheet, this is accomplished precisely by extrapolating the original operators $\Phi_\delta(X)$ to null infinity. In the Bondi coordinates $u,r,z,\bz$ of \eqref{udef} parametrizing the original coordinates $X^\mu$, we can express \eqref{PhiY} as
\be\label{tilPhi}
\til\Phi_\delta(Y) = \left(u+\frac{2r}{1+|z|^2}\right)^\delta\Phi_\delta(u,r,z,\bz)\,.
\ee
Let us study its large $r$ limit.

The large $r$ at fixed $u,z,\bz$ limit translates to the limit $Y^+\to0$ with $Y^-,Y^1,Y^2$ approaching the values shown in \eqref{Ybondi}. Denote the corresponding limit of $\til\Phi_\delta(Y)$ by
\be
\til{\O}_\delta(u,z,\bz) \vcentcolon= \lim_{r\to\infty}\til\Phi_\delta(Y)\,.
\ee
From \eqref{tilPhi}, we find that this is given by
\be
\til\O_\delta(u,z,\bz) = \left(\frac{2}{1+|z|^2}\right)^\delta\O_\delta(u,z,\bz)\,.
\ee
Here, $\O_\delta(u,z,\bz)$ denotes the large $r$ limit of $r^\delta\Phi_\delta(u,r,z,\bz)$ that was the extrapolation of $\Phi_\delta(X)$ to null infinity in the original dictionary. For a massless free scalar of weight $\delta=1$, it was given explicitly by \eqref{limrphi} in terms of creation and annihilation operators. 

One can now introduce celestial primaries \cite{Donnay:2022sdg}
\be\label{uMellin}
\O^\pm_{\delta,\Delta}(z,\bz) = \int_{-\infty}^\infty\d u\,(u\mp\im\epsilon)^{-\Delta}\,\O_\delta(u,z,\bz)
\ee
that are dual to the corresponding boost eigenstates of boundary weight $\Delta$. The $\im\epsilon$ prescription distinguishes between incoming $(-)$ and outgoing $(+)$ states as usual. The corresponding light-sheet celestial primaries are then easily read off:
\be
\til\O_{\delta,\Delta}^\pm(z,\bz) = \left(\frac{2}{1+|z|^2}\right)^\delta\,\O^\pm_{\delta,\Delta}(z,\bz)
\ee
which defines our new celestial extrapolate dictionary. Take the light-sheet $Y^+=0$ and projectivize it in rescalings of $Y^\mu$. Using the rescalings to set $Y^-=1$, the projectivization is seen to be a 2-sphere parametrized by $z=Y^1+\im Y^2$. Correlators of $\til\O^\pm_{\delta,\Delta}(z,\bz)$ inserted along this projectivized light-sheet will compute celestial correlators in the original conformal frame.

In this way, Hofman and Maldacena's conformal transformation has done something rather useful. Because of the extra factor of $(X^+)^\delta$, it has managed to automatically incorporate the factor of $r^\delta$ that we needed to multiply $\Phi_\delta(X)$ with to take its large $r$ limit. In this sense, it provides a more automated approach to the extrapolate dictionary. The prefactors of $2/(1+|z|^2)$ can of course be stripped off by hand. So that correlators of the $\til\Phi_\delta(Y)$ operators placed along the light-sheet compute correlators of the extrapolations of $r^\delta\Phi_\delta(X)$ to null infinity. We also note that while the quotient of the Einstein cylinder to a single Poincar\'e patch would map $\mathcal{I}^\pm$ to $Y^+\rightarrow 0^\pm$ from the uplift perspective we examined in section~\ref{sec:lift} the map~\eqref{eq:HM} is just a $\frac{\pi}{2}$ rotation in the $[-1,0]$ and $[3,4]$ planes~\cite{Hofman:2008ar}. The two null hyperplanes are on different Poincar\'e patches, and we can also keep track of the `missing generator' that compactifies the $Y^1,Y^2$ plane.


\bibliographystyle{jhep}
\bibliography{inv}

\end{document}